\documentclass[a4paper,12pt]{article}

\usepackage{cmap}					
\usepackage{mathtext} 				
\usepackage[T2A]{fontenc}			
\usepackage[utf8]{inputenc}			
\usepackage[english,russian,german]{babel}	

\usepackage{amsmath,amsfonts,amssymb,amsthm,mathtools} 
\usepackage{icomma} 

\usepackage{longtable}
\usepackage{multirow}

\mathtoolsset{showonlyrefs=false} 

\usepackage{graphicx}  
\graphicspath{{images/}{images2/}}  
\usepackage{wrapfig} 

\usepackage{array,tabularx,tabulary,booktabs} 
\usepackage{longtable}  
\usepackage{multirow} 

\theoremstyle{plain} 

\theoremstyle{definition} 

\theoremstyle{remark} 

\usepackage{etoolbox} 

\usepackage{geometry} 
\usepackage{fancyhdr} 
\usepackage{setspace} 
\usepackage{lastpage} 

\usepackage{soulutf8} 

\usepackage{hyperref}
\usepackage[usenames,dvipsnames,svgnames,table,rgb]{xcolor}
\hypersetup{				
    unicode=true,           
    pdftitle={Homogeneous functions of degree one and heat phenomena in potential fields.},   
    pdfauthor={Valentin K. Kochnev},      
    pdfsubject={Homogeneous functions},      
    pdfcreator={Valentin K. Kochnev}, 
    pdfproducer={Valentin K. Kochnev}, 
    pdfkeywords={Dark matter} {Hohenberg-Kohn functional} {Homogeneous functions}, 
    colorlinks=true,       	
    linkcolor=red,          
    citecolor=green,        
    filecolor=magenta,      
    urlcolor=cyan           
}

\usepackage{multicol} 
\usepackage{upgreek}
\usepackage{tikz}
\usepackage{pgfplots}

\pgfplotsset{compat=1.10}
\usepgfplotslibrary{groupplots}
\usepackage{csvsimple}
\usepackage{ dsfont }

\definecolor{plotgrid}{HTML}{ECECEC} 
\pgfplotsset{
	every axis legend/.append style={at={(0.5,-0.13)},anchor=north,legend cell align=left},
	tick label style={font=\tiny\scriptsize},
	label style={font=\scriptsize},
	legend style={font=\scriptsize},
	grid=both,
	minor tick num=2,
	major grid style={plotgrid},
	minor grid style={plotgrid},
	axis lines=left,
	legend style={draw=none},
	/pgf/number format/.cd,
	1000 sep={}
}

\pgfplotsset{
	/pgfplots/colormap={mycad}{rgb255(0cm)=(76,0,128) rgb255(2cm)=(0,0,255) rgb255(4cm)=(0,173,171) rgb255(6cm)=(32,205,0) rgb255(8cm)=(255,0,0) rgb255(10cm)=(128,0,0)}
}

\pgfplotsset{
	/pgfplots/colormap={matlab}{rgb255(0cm)=(0,0,128) rgb255(1cm)=(0,0,255) rgb255(3cm)=(0,255,255) rgb255(5cm)=(255,255,0) rgb255(7cm)=(255,0,0) rgb255(8cm)=(128,0,0)}
}

\renewcommand{\figurename}{Fig.}
\addto\captionsenglish{%
	\renewcommand{\figurename}{Fig.}%
}

\author{Valentin K Kochnev}
\title{One particle statistic.}

\begin{document} 

\begin{center}

Homogeneous functions of degree one and heat phenomena in potential fields.\\
\vspace{1.5ex}
 
Valentin K. Kochnev\\
 August 1, 2023\\
E-mail: valentine878@gmail.com
  \vspace{1.5ex}
  
\end{center}

The variational argument is presented to establish the attainability of homogeneity of degree one in the number of particles for any functional $F[n, f]$ that depends on both the state variable $f$ and the particle count $n$. Euler's integration of homogeneous functions applies to any such functional. This argument is employed to examine the heat equilibrium of a system containing an undefined and unconserved number of indistinguishable particles within each cell $h^3$ in the quantized phase-space of particle coordinates and momenta, with $h$ representing the Planck constant, including for the case of the smallest system of a single elementary volume. The system does not exchange particles with a reservoir, and the uncertainty in particle count is intrinsic to the system itself. The system is maintained at a constant temperature $T$ with a chemical potential denoted by $\mu$. The definition of chemical potential is based on variational principles related to homogeneous functions of degree one. The equilibrium particle density is analyzed in the presence of gravitational and electric fields characterized by a central $\frac{1}{r}$ reciprocally decaying potential, where the local density of potential field sources is provided for partition functions defined by the nature of the particles. A star is a point source of gravitation embedded in a rarefied ambient gas, where heat phenomena create a 'dark' illusion of additional mass presence. For an atomic core as a point source of an electric field in an electron gas, the study explores the temperature-dependent potential barrier in the electric field in atoms, where electron states correspond to particle states moving in the potential field with a barrier situated between a well and a valley. The homogeneous functionals for particle density energy and particle energy itself are discussed.

\section{The Objects of Variational Analysis}

Variational stationarity is the distinguishing appurtenance of physical systems and permits us to define exact equations, when we retain this property and abstract from all others, whether they be essential or accidental. Physical equations express the stationarity $\delta F = 0$ of a particular functional $F$, defined over a set of state variables or functions related to the physical system. The system's state is defined by those state variable values, which yield stationarity for the functional. The essence of the matter is that when contemplating a specific functional, we are always examining a \textit{class of equivalent functionals}, as deriving a functional from Euler's equations inherently involves ambiguity.

In physics, the most commonly encountered functionals are local functionals expressed in the form of integrals, as shown in equation \eqref{eq:1}:

\begin{equation}\label{eq:1}
F \big[n,\,f\big] = \idotsint_G W\big(n,\,x_1,\, \dots,\, x_n,\, f,\, f_{x_1},\, \dots,\, f_{x_n} \big) dx_1 \dots dx_n, 
\end{equation}

Here, $G$ represents the integration domain, and the argument of the functional is a multivariable function $f(x_1, \dots, x_n)$ and $f_{x_i}$ denote its first derivatives with respect to the variables $x_1, \dots, x_n$. The integrand $W$ also depends on the number of particles $n$, the parameter $n$ can be either a variable or non-variable parameter depending on context. It's emphasizing the intent to produce a homogeneous expression, while the homogenization of functions always introduces additional arguments. Every functional is defined by two factors: the set  $U$ of admissible elements $f$ on which it is given, and the integrand $W$ defining the law by which every element $f$ corresponds to a number, the value of the functional.

With certain assumptions regarding the differentiability of all functions involved, the expression for the variable functional leads to Euler's equation for $W$ in Eq. \eqref{eq:1}. This equation has the common form
$W_f - \frac{\partial}{\partial x_1} W_{f_{x_1}} - \frac{\partial}{\partial x_2} W_{f_{x_2}} - \cdots  - \frac{\partial}{\partial x_n} W_{f_{x_n}} = 0,$
where the subscripts denote derivatives with respect to corresponding variables of the integrand. Being a second-order equation in partial derivatives, its solution $f$ is sought either to take the given values at the boundary $\partial G$ of the integration domain or to satisfy transversality boundary conditions if variable boundary values are present or the boundary itself is variable. When $W$ depends on several variable functions or derivatives of higher order, the calculus of variations provides patterns for Euler's equations. These patterns are expressions for functional derivatives which, in turn, provide the formulation of physical laws in terms of variational principles.

The concept of \textit{class} reflects the abundance related to the fact that different functions and functionals may have the same necessary stationarity conditions -- different functionals may have the same functional derivatives. If we add to the expression $W$ under the integral in the functional \eqref{eq:1} the total differential of any function $\Phi(x_1,\, \dots,\, x_n,\, f)$, i.e. if we add an expression of the form
\begin{equation}\label{eq:2}
	\psi = \sum_i \frac{\partial \Phi}{\partial x_i} + \sum_i \frac{\partial \Phi}{\partial f} f_{x_i}
\end{equation}
to $W$, then the corresponding Euler's equation will not change, since the expression $\psi_f - \frac{\partial}{\partial x_1} \psi_{f_{x_1}} - \frac{\partial}{\partial x_2} \psi_{f_{x_2}} - \cdots  - \frac{\partial}{\partial x_n} \psi_{f_{x_n}}$ considered as a function of $x_i,\,f,\,f_{x_i}$ is identically equal to zero. Another sources of ambiguity when the functional is restored from Euler's equations are variables substitutions and variable transformations. Different integrals might yield Euler's equations locally equivalent up to variable substitutions and transformations.

Necessary assumptions about properties of functionals point to the set $\mathcal{F}$ of available functionals. It might be all local functionals with double continuously differentiable integrands by all its arguments while admissible functions have to be continuously differentiable if $W$ depends on first derivatives, double continuously differentiable if $W$ includes second derivatives, etc., for the applicability of analysis methods. The coincidence, up to a change or transformation of variables, of the Euler's equations corresponding to particular functionals in $\mathcal{F}$ defines the equivalence relation $\sim$ on $\mathcal{F}$. The $\mathcal{F}$ is divided into set $\mathcal{F}/\sim$ of disjoint equivalence classes of functionals with the same or equivalent Euler's equations. The class corresponding to a particular equation $\frac{\delta F}{\delta f}=0$ for the particular functional $F$ is denoted $\mathcal{F}/ \frac{\delta F}{\delta f}$ or $\mathcal{F}/ F$.

The initially defined functional \eqref{eq:1} selects the class $\mathcal{F}/ F$ of equivalent functionals from the factor-set $\mathcal{F}/\sim$. Once the quotient $\mathcal{F}/ F$ is determined, all its elements can be utilized to derive Euler's equations. Each equivalence class encompasses a continuum of infinitely many representatives. The patterns of Euler's equations are locally invariant under non-degenerate changes of variables. However, equivalence classes may have representatives whose existence is not immediately obvious but can be proven\footnote{For equivalence classes, linear operations are defined: $a_1 \cdot \big( \mathcal{F}/ F_1 \big)  + a_2\cdot \big( \mathcal{F}/ F_2 \big)$ is the equivalence class of the functional $a_1 \cdot F_1  + a_2\cdot F_2$, where $F_1$ and $F_2$ are any representatives of the classes  $\mathcal{F}/ F_1$ and $\mathcal{F}/ F_2$. A functional is identified with its equivalence class, and all reasoning for specific functionals, in particular, the attainability of a homogeneous structure, automatically applies to all representatives of the class.}. We will demonstrate that the functional $F[n,\,f]$ is equivalent to a homogeneous function of degree one in the number of particles, and to a homogeneous functional of degree one in the variable function $f$. By relying on the structure of homogeneous functions and homogeneous functionals, Euler's equation can be written even if the particular representative $F[n,\,f]$ was initially unknown.

The stationarity of functionals depending on the number of particles $F [n,\,f]$ is considered with constraints for the number of particles to ensure stationarity can occur. The equality constraints are incorporated using the Lagrange multiplier $\mu$, and the stationarity $\delta \{ F - \mu N \} = 0$ is considered for a linear combination of the functional of interest $F [n,\,f]$ with the functional of constraint\footnote{The latter is the left side of an equality $N[n,\,f] = N_{const}$ constraining the number $n$ in question, where the functional $N[n,\,f]$ may be defined as a local integral functional over the same domain $G$, or as an algebraic constraint (although the algebraic constraint not depending on $x_1,\, \dots,\, x_n,\, f,\, f_{x_1},\, \dots,\, f_{x_n}$ for simplicity, otherwise instead of a single Lagrange multiplier a dispertional relation would be in its place);} on the number of particles $N[n,\,f]$. It follows for the equilibrium value of $f$ that $\mu = \left( \frac{\delta F}{\delta N}\right)_{n,\,f}$ is a constant at all space points of the domain $G$. Here, the derivative on the right\footnote{It should be borne in mind that the sign in front of the \textit{non-zero} Lagrange multiplier for taking into account the equality condition is not important, nor is it important how to write the equation $N[n,\,f] = N_{const}$ or $N_{const}=N[n,\,f]$. What matters is the consideration of the \textit{linear combination} of the objective functional with the constraint functional. From the stationarity condition notation $\delta \{ F + \mu N \} = 0$, it would follow $\mu = -\left( \frac{\delta F}{\delta N}\right)_{n,\,f}$, and thus the notation $\delta \{ F - \mu N \} = 0$ is more often used in physics to get the chemical potential definition in the form $\mu = \left( \frac{\delta F}{\delta N}\right)_{n,\,f}$.} can depend on spatial coordinates and is equalized between all points of the system when the solution $f$ is achieved at a given constraint $N[n,\,f]=N_{const}$. When the functional $F$ signifies energy, the constant $\mu$ equalized between space points is termed the chemical potential.

\section{Particles in heat equilibrium}

In classical Thermodynamics\footnote{whose limitations we are going to overcome and completely drop in the next paragraph. This results in the extension of Thermodynamics laws to systems with any small number of particles, eradicating any assumptions that systems of particles can act purely quantum-mechanically at a positive temperature $T > 0\, K$, even if the number of particles is not \textit{large} enough;} the heat equilibrium is considered for a \textit{very large fixed} number of particles in the system. Joul's introduction of the mechanical equivalent of heat, i.e. universal coefficient $K$ for converting\footnote{$K=1$ if the heat and energy are expressed in the same units;} the work $A$ conducted on a system into heat $Q$ and expressing the latter in energy units $Q=KA$, led to the appearance of the differential of heat $\delta Q$ which is not a complete differential in arbitrary processes, the heat is not the function of the state and there is no direct possibility to compute $Q_2 - Q_1$ by integration when the system came from 1st state to 2nd\footnote{One has to pay attention to the fact that the integral functional of type \eqref{eq:1} cannot be written, and therefore, the equilibrium in the system cannot be studied by variational methods;}. To integrate the $\delta Q$ the integrating functional factor $\frac{1}{T(\theta)}$ is used to get the total differential $dS=\frac{\delta Q}{T}$, which yields the definition of the absolute temperature\footnote{The function $T(\theta)$ depends on the metric temperature $\theta$, measured by a thermometer in the system. The structure of the function reflects the construction of the device used to measure the temperature. If the temperature is measured in Celsius degrees by the calibrated extension or contraction of the volume $V$ of hydrogen gas in the thermometer, based on the Mendeleev-Clapeyron law $pV=NkT$, where $k$ is the Boltzmann constant, $p$ is the pressure, and $N$ is the number of particles, then the function is $T(\theta) = \theta + 273.15\,K$;} of the system $T$ and the function of state $S$, the entropy. The energy of the system $U(S,\,V,\,N)$ depends on a new variable $S$, involves the heat and is the function of the state, the equilibrium of the system is defined by $\delta \left\{U - \mu N\right\} = 0$, that is for a given number of particles $N=N_{const}$. It is common instead of the explicit Lagrange multiplier to apply the stationarity statement to the partial value of energy $U_m$ related to one mole of particles or another unit, while still writing $\delta U = 0$ omitting index $m$ and considering the chemical potential $\mu$ independently as $\mu = \left(\frac{\partial U}{\partial N}\right)_{S,\,V,\,N=N_{const}}$. Then the chemical potential can be used for open systems as $dU = TdS-pdV+\mu dN$ so that the united system with the reservoir has a fixed number of particles. Both notations are equivalent in the sense that both yield the same values of intensive variables when stationarity is achieved, but for partial thermodynamic potential, all other extensive properties must also be treated as related to one mole. Negated\footnote{the negated Legendre transform $H=-U^*$ instead of standard transform $U^*$ is used for $V$, because the pressure is $p=-\left(\frac{\partial U}{\partial V}\right)_{S,\,N}$} Legendre transform by the second variable is used to get the equivalent representation of the internal energy $U$, as the enthalpy $H(S,\,p,\,N)$ which depends on the intensive variable the pressure $p$ instead of the extensive variable the volume $V$. Legendre transforms by the first variables $S$ produce the Helmholtz energy $F$ from the internal energy $U$ and the Gibbs energy from the enthalpy $H$, both depend on the intensive variable the temperature $T$ instead of the extensive variable the entropy $S$. All thermodynamic potentials $U=TS-pV+\mu N,\,H = TS + \mu N,\,F = -pV + \mu N,\,G = \mu N$ are extensive functions of state and can be used for the equilibrium definition with the constraint on the number of particles $N$ in the system of interest. Since these are equivalent representations of energy, and the Legendre transformation does not change the number of particles in the system, then\footnote{Compare this to the fact that the shift of the functional \eqref{eq:1} with Eq. \eqref{eq:2}, producing the functional $F_1$ from $F$, does not change the value of the Lagrange multiplier when the variation $\delta \left\{F_1 - \mu N\right\} = 0$ is used instead of $\delta \left\{F - \mu N\right\} = 0$. Note that Eq. \eqref{eq:2} shifts only the additive constant added to the integral and does not affect the difference;}
\begin{equation}\label{eq:3}
	\mu = \bigg(\frac{\partial U}{\partial N}\bigg)_{S,\,V} = \bigg(\frac{\partial H}{\partial N}\bigg)_{S,\,p}=\bigg(\frac{\partial F}{\partial N}\bigg)_{T,\,V}=\bigg(\frac{\partial G}{\partial N}\bigg)_{T,\,p}.
\end{equation}
The chemical potential $\mu$ is a thermodynamic potential related to one particle\footnote{The ISO (International Standards Organization) recommends that the chemical potential is defined as a partial value of the Gibbs energy. Of course, this recommendation should not lead to misconceptions regarding the physical nature of chemical potentials. The reason for this recommendation is that the Gibbs energy at the equilibrium state of the system $G=\sum \mu_i N_i$ is an explicit homogeneous function of degree one for the number of particles. Other thermodynamic potentials are also homogeneous of degree one for the number of particles at the equilibrium state of the system, but implicitly so, due to the extensivity of the volume and the entropy. }.

The extensivity of the internal energy $U$ and its transforms relies in classical Thermodynamics on a thermodynamically large number of particles in the system so that the characteristic length of interactions between particles is negligible compared to the geometric dimensions of the particle system\footnote{when the energy of interactions between particles is described by the power law $\frac{1}{R^\alpha}$ where $R$ stands for a distance between two particles and commonly $\alpha = 1 \div 6$, interactions take place for any distance, but for large enough $R$ the energy of interaction is infinitesimal, while $R$ is still negligible compared to system geometric dimensions;}. This provides the independence of the thermodynamical expressions and laws, which essentially rely on the structure of homogeneous functions of degree one on the number of particles, from the nature of particles and interactions between the particles, the independence of common forms on the state equation of a particular system\footnote{the attainability of the homogeneity of degree one on the number of particles is pointed out to be the property of the equivalence class of related functionals and no longer relies on the number of particles in the system;}.

For the state of ideal \textit{homogeneous} gas, the chemical potential \eqref{eq:3} is
\begin{equation}\label{eq:4}
	\mu = \mu_0 + kT \ln n,
\end{equation}
where $\mu_0$ is a constant at given temperature, $k$ is the Boltzman constant, and the $n = \frac{N}{V}$ is a number density of particles in the gas\footnote{the expression \eqref{eq:4} is given in statistical physics for a monatomic gas, using the Sackur-Tetrode equation, and then the constant $\mu_0$ has an \textit{absolute} value, as $\mu_0=3kT \ln \frac{h}{\sqrt{2\pi m kT}}$ where $h$ is a Planck constant and $m$ is a mass of one particle. Except for the particular expression for the $\mu_0$ the equation itself is not restricted to monoatomic gases;}. This equation also is given for the component pressure\footnote{for the volume at constant temperature $V=\big(\frac{\partial G}{\partial p}\big)_{T,\,N}$ and $V=\frac{NkT}{p}$ the integration for $p$ yields the free energy $(G-G_0)$ that can be released in the form of work during the transition of the system from $p$ to $p_0$ as $G = G_0 + NkT \ln \frac{p}{p_0}$, thus the potential related to one particle is $\mu = \mu_{0,\,p} + kT \ln \frac{p}{p_0}$. To obtain an unitless quantity $\frac{p}{p_0}$ under the logarithm sign, the standard state of matter is selected at $p_0 = 1\,\left[p\right]$, and the constant $\mu_{0,\,p}$ gets the \textit{absolute} value of the standard potential at a given temperature, i.e. potential $\mu$ for the standard state of matter.} $p$ in the ideal gas $\mu = \mu_{0,\,p} + kT \ln p$, and for the component mole fraction $x$ in an ideal solution $\mu = \mu_{0,\,x} + kT \ln x$, or the component mole concentration \footnote{an \textit{ideal} solution is the solution satisfying the Raoult's law $p=p_0 x$, where $p$ is the partial pressure of the component in the gaseous mixture above the solution, $p_0$ is the equilibrium vapour pressure of the pure component, $x$ is the mole fraction of the component in the liquid or solid solution, thus $\mu = \mu_{0,\,p} + kT \ln p = \mu_{0,\,p} + kT \ln p_0 x = \mu_{0,\,x} + kT \ln x$, for an extremely dilute ideal solution, the mole fraction $x$ can be expressed in terms of the mole concentration $c$, giving another expression $\mu = \mu_{0,\,c} + kT \ln c$;} $c$ in an ideal solution $\mu = \mu_{0,\,c} + kT \ln c$, where corresponding constants $\mu_{0,\,p},\,\mu_{0,\,x},\,\mu_{0,\,c}$ depend on temperature, but not depend on pressure, mole fraction or concentration. These expressions of chemical potential differ in the choice of the standard state of the matter to obtain an unitless quantity under the logarithm sign.

The chemical potentials are interpreted as thermodynamically determined potential energies, related to heat properties of the system\footnote{in the absence of chemical reactions, otherwise, related to both heat and chemical properties;}, but in all other aspects behaving as mechanical potentials.    
 
 \renewcommand{\figurename}{Fig.}
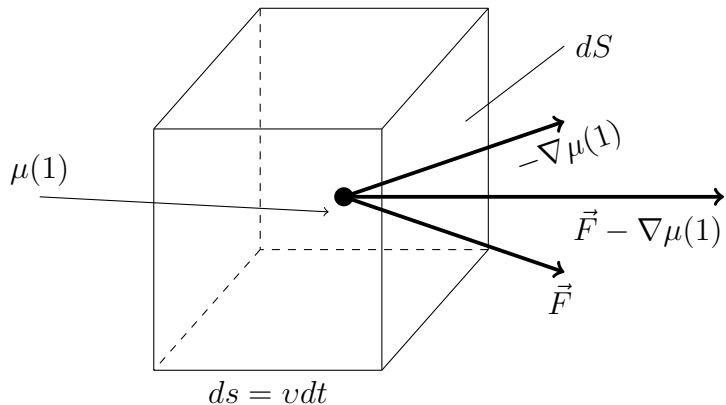
\begin{figure}
	\begin{center}
		\begin{tikzpicture}[domain=0:6]
			\draw[-] (-0.6,0.0) -- (-2.0, -1.6);
			\draw[-] (-0.6,3.2) -- (-2.0, 1.6);	
			\draw[-] (-2.0, -1.6) -- (-2.0, 1.6);
			\draw[-] (-0.6,3.2) -- (-0.6,0.0);
	\draw[dashed] (-3.6,0.0) -- (-5.0, -1.6);
    \draw[-] (-3.6,3.2) -- (-5.0, 1.6);	
   \draw[-] (-5.0, -1.6) -- (-5.0, 1.6);
   \draw[dashed] (-3.6,3.2) -- (-3.6,0.0);
		\draw[dashed] (-0.6,0.0) -- (-3.6, 0.0);
   		\draw[-] (-0.6,3.2) -- (-3.6, 3.2);	
		\draw[-] (-2.0, -1.6) -- (-5.0, -1.6)node[pos=0.5,below] {$ds=\upsilon dt$};
			\draw[-] (-2.0,1.6) -- (-5.0,1.6);
\fill (-2.5,0.7) circle (3.6pt);
\draw[->, line width=0.5mm] (-2.5,0.7) -- (2.5,0.7) node[pos=0.8,below] {$\vec{F}- \nabla \mu(1)$};
\draw[->, line width=0.5mm] (-2.5,0.7) -- (0.4,1.7) node[pos=0.98,sloped,below] {$- \nabla \mu(1)$};
\draw[->, line width=0.5mm] (-2.5,0.7) -- (0.4,-0.3) node[pos=0.98,below] {$\vec{F}$};
\draw[] (-0.9,1.7) -- (0.4,2.7) node[right] {$dS$};
\draw[->] (-6.5,0.7)--(-2.7,0.5) node[above,pos=0] {$\mu(1)$};
		\end{tikzpicture}
		\caption{\label{fig:dV} Forces on a particle in a volume element $dV$.}
	\end{center}
\end{figure}

The equation \eqref{eq:4} can be applied to the inhomogeneous ideal gas if considering the system <<in small>>, that is within the small enough volume element $dV$ where the inhomogeneity is neglectable. Then it is a function of spatial coordinates $\mu(1) = \mu_0 + kT \ln n(1)$, where the number density is expressed for particles within a volume element $n(1) = \frac{dN}{dV}$ and the gradients of chemical potentials in the system induce mechanical forces acting on particles\footnote{the standard assumption about a volume element $dV$ is it is large enough to contain a thermodynamically large number of particles $dN$ to provide the extensivity of thermodynamic potentials for these particles and the expression $\mu(1) = \mu_0 + kT \ln n(1)$ for  potential per particle; this assumption will be dropped further and the element $dV$ may be as small as dictated by the size of elementary cell $h^3$ for a Plank constant $h$;}, Fig. \eqref{fig:dV}. The differentiating of a local chemical potential $\mu(1)$ yields the vector $- \nabla \mu(1) = -kT \frac{1}{n(1)} \nabla n(1)$ accelerating the particles in the direction towards lower number density and acting together with the force $\vec{F}$ generated by some force field. The flux of particles through an area element $dS$ perpendicularly situated to the force on particles can be expressed by mechanical laws. For the density with zero velocity (non-moving gas or solution), when the particles are not transferred by the motion of the density as a whole the acceleration of particles by forces within the element $dV$ is $\vec{a} = \frac{1}{m}\vec{F}-\frac{kT}{m}\frac{1}{n(1)} \nabla n(1)$. The particles in the density do not accumulate velocity, but the density itself can change, the speed after acceleration under the action of forces in the volume element and the corresponding displacement of particles have meaning exactly within the differential volume. The velocity gained by particle for the time $dt$ is $\vec{\upsilon}=\frac{1}{m}\left(\vec{F}-\frac{kT}{n(1)} \nabla n(1)\right)dt$ and the displacement of the particle with this speed\footnote{the factor $\frac{1}{2}$ due to the finite formula $s=\frac{at^2}{2}$ has not occurred, because we do not pass to finite values. Consider the velocity for the time $dt$, as $\upsilon = \int_{0}^{dt} a\,dt = a dt$, and the distance for $dt$, as $ds=\upsilon dt$;} for the time $dt$ is $ds=\frac{1}{m}\left|\vec{F}-\frac{kT}{n(1)} \nabla n(1)\right|dt^2$. The thickness of the volume $dV$ perpendicular to the direction of particle displacement is equal to the displacement $ds$ for the time $dt$, Fig. \eqref{fig:dV}. The volume is $dV = ds \, dS$ contains $dN = \frac{n(1)}{m}\left|\vec{F}-\frac{kT}{n(1)} \nabla n(1)\right|dt^2\cdot dS$ particles. All these particles pass through the surface $dS$ for the time $dt$ and the flux of particles is $d\vec{j}=\frac{dN}{dt\cdot dS} \,\vec{e}=\frac{n(1)}{m}\left(\vec{F}-\frac{kT}{n(1)} \nabla n(1)\right)dt$, where $\vec{e}$ is the orth of $\left(\vec{F}-\frac{kT}{n(1)} \nabla n(1)\right)$. The flux of particles for the unit of time at an arbitrary point
\begin{equation}\label{eq:flux}
	\vec{j} = \frac{n(1)}{m} \vec{F} - \frac{kT}{m} \nabla n(1)
\end{equation}
consist of the contribution of the force field and the contribution of heat processes. If the temperature is zero, all flux is related to the force field. If there is no force field, the flux does satisfy Fick's law\footnote{indicating that this law is applicable within the limits of applicability of the ideal gas equations of state;}
\begin{equation}\label{eq:Fick}
	\vec{j} = - D \,\nabla n(1)
\end{equation}
with the temperature-dependent coefficient\footnote{Of course, in many real situations, instead of the mass of particles, there should be some of their effective mass. For example, in solution the particles are solvated, etc.} of diffusion $D=\frac{kT}{m}$.

The movement of particles, unlike the case of mechanical material points, must satisfy the continuity of the medium $\frac{\partial n}{\partial t} +  \nabla \cdot \vec{j} = 0$, which expresses that the gas of particles or liquid does not show discontinuities during the perpetual movement of the particles of which they consist. It reads\footnote{$\nabla \cdot \vec{j} = \frac{1}{m} \nabla \cdot (n \vec{F})-\frac{kT}{m} \nabla \cdot \nabla n = \frac{1}{m} \nabla n \cdot \vec{F} + \frac{1}{m} n \nabla \cdot \vec{F} - \frac{kT}{m} \triangle n$;} $\frac{\partial n}{\partial t} + \frac{1}{m} \nabla n \cdot \vec{F} + \frac{1}{m} n \nabla \cdot \vec{F} - \frac{kT}{m} \triangle n=0$. When there is no force $\vec{F}$ the equation is the Diffusion Equation. For potential fields the force $\vec{F}$ is defined by the gradient of the potential. For gravitation, the force is $\vec{F} = -m \nabla \varphi$ for the gravitational potential $\varphi$, which can be found by solving the system of equations\footnote{here in the right part of the Poisson equation, the mass density $m \cdot n$ is produced from the number density of particles $n$;} 
\begin{equation}\label{eq:thermal_effects}
		\begin{cases}
			\frac{\partial n}{\partial t} -  \nabla n \cdot \nabla \varphi - n \triangle \varphi - \frac{kT}{m} \triangle n=0
			\\
			\triangle \varphi = 4\pi Gm \cdot n
	\end{cases}
\end{equation}
for functions $n(t,\,x,\,y,\,z)$ and $\varphi(t,\,x,\,y,\,z)$ with consistent boundary and initial conditions.

The most important significance of the system \eqref{eq:thermal_effects} is that it demonstrates that within the broadest scope of the ideal gas laws and even for ideal systems themselves, potential fields depend on heat phenomena at any positive absolute temperature.

The heat equilibrium of particles in space is characterized by the equalization of a chemical potential $\mu$ for these particles at all points in space\footnote{This is the fundamental property of Lagrange multipliers. However, the term 'chemical potential' should be used carefully. It can be a full potential with contributions from several potential fields. The potential $\mu = \mu_0 + kT \ln n$ is a Lagrange multiplier for the Gibbs energy of a \textit{homogeneous} gas at constant pressure and temperature, while it is minimized $\delta \left\{G -\mu N\right\} = 0$ for a given number of particles. But for an \textit{inhomogeneous} gas, described by Eqs. \eqref{eq:thermal_effects}, the $\mu(1) = \mu_0 + kT \ln n(1)$ is not a Lagrange multiplier. Instead, one might write down the functional $F[n(t,\,x,\,y,\,z),\,\varphi(t,\,x,\,y,\,z)]$ for which Eqs. \eqref{eq:thermal_effects} are Euler equations -- and the functional might be a sum of thermodynamic energy and energy of potential field -- for the direct solution of the system \eqref{eq:thermal_effects} with a constraint on the number of particles $\delta \left\{F -\eta N\right\} = 0$ with an equalized Lagrange multiplier $\eta$ representing the full potential, while using the potential $\mu(1) = \mu_0 + kT \ln n(1)$ in <<small>>;}. Thermodynamically, the moving force driving particle density flow and redistribution is the difference in chemical potential $\mu$. In a state of heat equilibrium, when no moving forces are present, $\mu$ is equalized between all points of space, including at infinitely distanced locations\footnote{Even if there are no particles there -- the Lagrange multiplier is the property of the system as a whole, and not a property of medium points -- otherwise particles would flow to infinity;}. The potential $\varphi$ of the force field for particles is an auxiliary variable, although its presence affects the value of the chemical potential $\mu$, and vice versa. For the state of ideal homogeneous gas it is \eqref{eq:4}. For the inhomogeneous ideal gas \eqref{eq:thermal_effects}, it is full potential consisting of contributions of all potential fields, including the local chemical potential $\mu(1)$ expressed by logarithm.  For homogeneity of gas in phase-space of particles' spatial coordinates and momenta rather than spatial space, the phenomenon is manifested the most straightforwardly -- in quantum statistics expressed by Fermi-Dirac and Bose-Einstein integrals the chemical potential $\mu$ is a constant throughout space, being a part of full potential which varies between points -- this is due to the independence of potentials on particle momentum. While the forces acting on particles vanish at distanced points from all sources of potential fields, the chemical potential $\mu$ remains constant. The value of chemical potential for the particles can be found by Eq. \eqref{eq:4} for a spatially homogeneous ideal gas, or generally by a more sophisticated universal method below, applicable for any non-ideal system, based on the structure of homogeneous functions.

Near the point of asymptotic boundary condition $\frac{1}{r}$, the number density can be very large, and the first equation of \eqref{eq:thermal_effects} is applicable only far enough from asymptotic centers. The heat properties of any density of particles can be accounted for directly on the right side of the Poisson equation with a Fermi-Dirac integral or Bose-Einstein integral. 

For a consideration of the system <<in small>>, we will need to consider the heat equilibrium of an indefinite and unconserved number of particles of non-zero rest mass within a single cell $h^3$ of quantized phase-space of coordinates and momenta of a particle, where $h$ is the Planck constant. The number of particles in such a cell is a quantum observable for the number of particles. If the cell were related to the vicinity of particular spatial coordinates $x,\,y,\,z$ and a specific value of momentum $\vec{p}$, then due to uncertainties $\triangle x \cdot \triangle p_x \geq \frac{h}{4 \pi},\,\triangle y \cdot \triangle p_y \geq \frac{h}{4 \pi},\,\triangle z \cdot \triangle p_z \geq \frac{h}{4 \pi}$, nothing can be said about the exact population of a particular cell $h^3$, nor can it be normed to a specific number of particles\footnote{For the same reason, it is generally not possible to normalize a local spatial density of particles based on quantum statistics -- it is not possible to normalize the Fermi-Dirac integral or Bose-Einstein integral, although such normalization might be seen in the literature;\label{footnote_23}}. 

The definition of the chemical potential for a system with an undefined and unconserved number of particles follows as a necessary condition for the variational stationarity of a sum of physical values\footnote{For a system of an undefined number of particles, the differentiation by the number of particles is not defined. In classical Thermodynamics, the indefiniteness of the number of particles is considered in terms of the exchange of particles between the system and an outer reservoir of particles so that the united system with a reservoir has a fixed thermodynamically \textit{large} number of particles $N$, all extensive variables attain the homogeneity of degree one in the number of interacting particles, and the coefficients of the homogeneous forms are partial values -- derivatives by the number of particles at a specific fixed $N$. As well as the attainability of the homogeneity of degree one in the number of interacting particles no longer relies on the size of $N$, the definition of the chemical potential is considered here in terms of differentiability of a functional $F\left[n,\, f\right]$ -- for any possible $n$ there is a corresponding value of $\mu$ -- and the explicit augmentation of a system of interest with a large reservoir is no longer necessary. The equilibrium is defined by the canonical ensemble while yielding the same relations for probabilities, as the grand-canonical ensemble for an augmented system;}.

\section{The definition of chemical potentials}

General statements for a system of particles can be established using only the linear properties of space, the addition of elements, and the multiplication of the elements by a number, which are properties of arithmetic space not equipped with a metric. This provides a general definition of the chemical potential below.

A variable number $F[n, \varphi]$ called a functional depends on a variable function $\varphi$ and a parameter $n$. The linear combination
\begin{equation}\label{eq:combination}
    F[a_i, n_i, \varphi_i] = \sum_i a_i F_i[n_i, \varphi_i]
\end{equation} 
of such functionals\footnote{One can think that $F_i[n_i, \varphi_i]$ is the Gibbs energy of the $i$-th part of the system, while $\varphi_i$ is a state function of the part. However, for the definition, it does not matter at all; it might be the volume of the $i$-th part of the system, the energy of $i$-th formal copy of the system as a whole or the energy of $i$-th particles, etc.;} also depends on the coefficients $a_i$. The stationarity condition for this linear combination is investigated subject to requirements of fixed products\footnote{Kinetic energy of material points $T = \sum_i \frac{m_i}{2}(\dot{x_i}^2+\dot{y_i}^2+\dot{z_i}^2)$ and potential energy of interacting masses $U=-G\sum_{i<j}\frac{m_i m_j}{r_{ij}}$ are recognized by such a common pattern. Wherein fixed values $m_i$ mean constant masses of material points and no redistribution of mass between bodies;} $N_i[a_i, n_i] = n_i a_i$. The number of terms in the combination is either fixed or undefined but required to be equal to the number of fixed values $n_i a_i$. Variational arguments following do not depend on the number of terms and are valid for any undefined number of them. Thus, the conclusions of the variation calculation are valid for an undefined number of subsystems or an undefined number of particles in the system.

Taking into account the constraints using Lagrange multipliers $\mu_i$, the stationary condition reads
\begin{equation}\label{eq:argument}
	\delta \biggl\{ F[ \, a_i,\,n_i, \varphi_i \,] - \big( \mu_1\,N_1[a_1,\,n_1\,] + \dots +\mu_k\,N_k[a_k,\,n_k\,]+ \dots \big) \biggr\}=0.
\end{equation}
Differentiation gives
\begin{multline}
	\sum_i\,\delta a_i \Bigg[ \Bigg( \dfrac{\partial F}{\partial a_i} \Bigg)_{a_j, n, \varphi} - \mu_i\Bigg( \dfrac{\partial N_i}{\partial a_i} \Bigg)_{a_j, n, \varphi\,} \Bigg]	+ \\ \sum_i\,\delta n_i \Bigg[ \Bigg( \dfrac{\partial F}{\partial n_i} \Bigg)_{a, n_j, \varphi} - \mu_i\Bigg( \dfrac{\partial N_i}{\partial n_i} \Bigg)_{a, n_j, \varphi\,} \Bigg] + \\ \sum_i\,\delta \varphi_i \, a_i \dfrac{\delta F_i}{\delta \varphi_i} = 0.
\end{multline}
Each term must be zero for arbitrary independent variations of coefficients $\delta a_i$, of parameters $\delta n_i$, and of states $\delta \varphi_i$. For coefficients after $\delta a_i$ it follows that 
\begin{equation}\label{eq:homogeneity}
F_i = \mu_i\,n_i
\end{equation}
While for a coefficients after $\delta n_i$ it gives $a_i\,\dfrac{\partial F_i}{\partial n_i} = \mu_i\,a_i$, i.e. 
\begin{equation}\label{eq:chem_potentials}
\dfrac{\partial F_i}{\partial n_i} = \mu_i.
\end{equation}
These simple statements for the stationarity of the linear combination are the essence of the assay. Equation \eqref{eq:chem_potentials} provides the definition for chemical potentials in the system of particles when parameters $n_i$ have a meaning of the number of particles or are related to the number of particles, such as states populations, masses, and moles. Equation \eqref{eq:homogeneity} expresses that physical values related to the number of interacting particles can always be written as the number of particles times the partial value, given the equalized chemical potentials. It does not matter how the spatial extent of interactions between particles and the characteristic size of the system relate to each other, or as if there were no interactions at all\footnote{Applying the variational argument \eqref{eq:argument} to the equivalence classes $\mathcal{F}/ F_i$ of functionals $F_i$, we obtain that each equivalence class $\mathcal{F}/ F_i$ has a representative homogeneous of the first degree in the number of particles.}.

For a particular case of a conserved number of particles $N$, the combination of equations \eqref{eq:homogeneity} and \eqref{eq:chem_potentials} yields the definition of the chemical potential for a system with a conserved number of particles as a derivative with respect to $N$: if the particles of a given $i$-th type with a potential $\mu_i$ are distributed among $k$ parts of the system so that $\sum_{k} n_k = N$, and the extensive value $F_i$ is summed for the $i$-th particle type as $F_i = \sum_{k} F_k$, then $F = \sum_{k} \mu_i n_k = \mu_i \sum_{k} n_k = \mu_i N$, and $\mu_i = \left( \frac{\partial F_i}{\partial N} \right)_{a_j, \varphi}$, where the partial derivative is computed with fixed other variables.

For coefficients after $\delta \varphi_i$, the equations $\frac{\delta F_i}{\delta \varphi_i} = 0$ are equations of mechanics, field theory, etc. The computation of functional derivatives $\frac{\delta F_i}{\delta \varphi_i}$ for these equations is the essence of classical calculus of variations.

The equation \eqref{eq:chem_potentials} expresses the equalization of chemical potentials between all points of space. The left part of it can depend on spatial coordinates, but the right part is a constant Lagrange multiplier. A particular case of this equation is known for the analysis of natural orbitals in quantum mechanics, as Parr's principle of equal orbital\footnote{See equation (57) in the section ’PRINCIPLE OF EQUAL ORBITAL ELECTRONEGATIVITIES’ in \cite{Parr1978};} electronegativities \cite{Parr1978, Gilbert1975}. Similarly, a particular case of Eq. \eqref{eq:homogeneity} along with the equation \eqref{eq:chem_potentials} was also derived for natural orbitals \cite{Kochnev2017, Kochnev2017_1}.

\section{Homogeneous functions of degree one}
The linear combination gets simplified to 
\begin{equation}\label{eq:13}
F[ \, a_i,\,n_i, \varphi_i \,] = \sum_i\,a_i\,\mu_i\,n_i,
\end{equation}
when the stationarity necessary conditions are met. This conclusion does not depend on the choice of variables $n_i$. Such forms, known as homogeneous functions of degree one, express extensive variables of state in equilibrium thermodynamics. Commonly, $n_i$ represents the number of particles or proportional mass, and derivatives $\mu_i$ according to Eq. \eqref{eq:chem_potentials} are chemical potentials or other intensive partial values. Thermodynamic potentials are expressed in this way.

The variational argument \eqref{eq:13} extends Euler's theorem on the structure of homogeneous functions. While Euler's theorem defines a form \eqref{eq:13} for any homogeneous function of degree one, Eqs. \eqref{eq:homogeneity}--\eqref{eq:chem_potentials} express the necessity to transform any function for particles into a homogeneous function of degree one when the required stationarity conditions \eqref{eq:homogeneity}--\eqref{eq:chem_potentials} are met\footnote{That is, to take its homogeneous function of degree one version from the corresponding equivalence class.}.

The linear combination \eqref{eq:combination} corresponds to the state of an ideal gas, consisting of systems of interest containing interacting particles, such as molecules in the gas. Homogeneous functions of degree one in terms of the number of particles then follow from the most general variational stationarity of physical values taken together in the linear combination. A well-defined physical value $F[n, \varphi]$ is either defined as a homogeneous expression of degree one\footnote{For example, total energy $H = T + U$ and the Lagrange function $L = T - U$ characterize a mechanical system of material points. It is instructive that the functions $H$ and $L$ for interacting material points are defined as homogeneous functions of degree one with respect to masses $m_i$, that is $H = \sum_i \frac{\partial H}{\partial m_i} m_i$ and $L = \sum_i \frac{\partial L}{\partial m_i} m_i$. The action $S = \int_{t_0}^{t_1} L dt$ is also a homogeneous function of degree one, given by $S = \sum_i \left( \int_{t_0}^{t_1} \frac{\partial L}{\partial m_i} dt \right) \cdot m_i$;\label{footnote_29}} or, if it is generally non-homogeneous, it always becomes a homogeneous expression of degree one when stationarity requirements are met for an ideal gas.

If $\rho$ in the functional $F[n, \rho]$ represents a spatially-based density of $n$ particles\footnote{As mentioned in footnote \eqref{footnote_23}, it is necessary to clearly distinguish between the spatially-based density of particles and the phase-space-based density of particles. The former is normed to a number of particles in space, while the latter cannot be normed to a number of particles due to uncertainty relations;}, then Eqs. \eqref{eq:homogeneity}-\eqref{eq:13} lead to a \textit{homogeneous integral expression}\footnote{The functional $F[n, \rho]$ is homogeneous of degree $k$ in $\rho$ if $\int \frac{\delta F[n, \rho]}{\delta \rho} \rho \,dV = k F[n, \rho]$;} for the functional $F[n, \rho]$. The spatially-based density $\rho$ is always normed to the number of particles $n$ in the system, such as $n = \int \rho dV$. Equation \eqref{eq:homogeneity} then becomes $F[n, \rho] = \mu \int \rho dV$. Differentiating the latter gives $\frac{\delta F}{\delta \rho} = \mu$, and introducing a constant $\mu$ under the integral sign yields\footnote{Thus, assuming the functional $F[n, \rho]$ is \textit{well-defined} but its expression is not known, we found the expression as a local integral functional homogeneous of degree one in $\rho$; } $F[n, \rho] = \int \frac{\delta F}{\delta \rho} \rho \,dV$. If $\rho$ represents the phase-space-based density of particles, and it is \textit{summable} $||\rho||_{L_1}=\int \rho dV < \infty$, then the same expression as a local integral functional homogeneous of degree one in $\rho$ follows, such as $n = d ||\rho||_{L_1}$, where the coefficient $d>0$, the equation \eqref{eq:homogeneity} then becomes $F[n, \rho] = \mu d \int \rho dV$. Differentiating the latter gives $\frac{\delta F}{\delta \rho} = \mu d$, and introducing a constant $\mu d$ under the integral sign yields $F[n, \rho] = \int \frac{\delta F}{\delta \rho} \rho \,dV$. The coefficient $d>0$, which characterizes the distribution of particles by momenta at a given temperature, does not introduce any indefiniteness, because the Lagrange multiplier $\mu$ is still undefined. It follows that the normalization equation to define an unknown constant is indifferent to defining $\mu$ or $\mu d$, so if the spatially-based density of particles is not known, the known phase-space-based density of particles can be exploited to compute the value of functional $F[n, \rho]$. The homogeneous functional $F[n, \rho] = \int \mu \rho \,dV$ itself is defined up to the equivalence expressed by Eq. \eqref{eq:2}. Shifting the integrand by a $\psi = \varphi_0 \rho$, where $\varphi_0$ is a zero-valued at infinity solution of a particular Poisson equation $\Delta \varphi_0 = \rho$ gives $F[n, \rho] = \int \varphi \rho \,dV$, where $\varphi$ is potential equal to $\mu$ at intinity, and $\frac{\delta F}{\delta \rho} = \varphi$ \footnote{the equivalence-based flexibility of changing the expression for a homogeneous functional is in a direct analogy with a well-known theorem for homogeneous functions: any homogeneous function of degree one can be presented $f(x_1,\,x_2,\dots,\,x_n)=\phi(x_1,\,x_2,\dots,\,x_n) \cdot h(\phi_1(x_1,\,x_2,\dots,\,x_n),\,\phi_2(x_1,\,x_2,\dots,\,x_n),\dots,\, \phi_{n-1}(x_1,\,x_2,\dots,\,x_n))$, where $h(t_1,\,t_2,\dots,\,t_{n-1})$ is some suitable function of $(n-1)$ variables, $\phi$ is a fixed homogeneous function of degree one, and $\phi_1(x_1,\,x_2,\dots,\,x_n),\,\phi_2(x_1,\,x_2,\dots,\,x_n),\dots,\, \phi_{n-1}(x_1,\,x_2,\dots,\,x_n)$ are fixed functionally independent homogeneous functions of zero degree. For a fixed choice of functions $\phi,\,\phi_1,\,\phi_2,\dots,\, \phi_{n-1}$, this representation specifies a one-to-one correspondence between homogeneous functions $f$ of degree one of $n$ variables and functions $h$ of $(n-1)$ variables. The correspondence itself is defined up to the equivalence relation -- for a fixed left side $f$ many different right sides $\phi \cdot h(\phi_1,\,\phi_2,\dots,\,\phi_{n-1})$ can be used.}.

Equating the two forms yields a general normalization equation
\begin{equation}\label{eq:normalization}
	 \mu n = \int \frac{\delta F}{\delta \rho} \rho \,dV
\end{equation}
which is applicable to a phase-space-based density of particles. In this case, the integral $\int \rho dV$ has no sense as the number of particles due to the uncertainty relations. The energy of density $\rho$ in the \textit{outer} potential $\frac{\delta F}{\delta \rho}$ is the energy of $n$ particles in the medium with the chemical potential $\mu$.

In the context of heat equilibrium, the potential $\varphi = \frac{\delta F}{\delta \rho}$ cannot have an arbitrary reference point as it could in a pure mechanical context. The reference point of the potential $\varphi$ does not impact any difference in potential between points in space, nor does it affect the force field $-\nabla \varphi$. Similarly, the reference point of the chemical potential $\mu$ does not influence any difference in $\mu$ between points in space. However, the references of potentials $\varphi$ and $\mu$ have to be consistent according to Eq. \eqref{eq:normalization}.

At distant points from all sources of the field, the force $\vec{F}$ vanishes, but the potential $\mu$ remains constant. The consistency \eqref{eq:normalization} is achieved if the reference point of the potential $\varphi$ is consistent with the value of the chemical potential $\mu$ so that when there is no field $\vec{F}$, the potential energy of a particle is only thermodynamically determined potential energy\footnote{It follows that the absolute value of the chemical potential restricts the field potential $\varphi$ to an absolute value in Eqs. \eqref{eq:16}-\eqref{eq:17} below. In particular, for the gravitational potential around the point source, it is necessary to choose $\varphi(+\infty)=-\frac{\mu}{m}$, and for the electrostatic potential around the atomic core, it is necessary to choose $\varphi(+\infty)=-\frac{\mu}{q}$, that is $\varphi(+\infty)=\mu$ for electrons when expressed in atomic units, which provides the vanishing of particles' density at infinitely distanced points;\label{footnote_34}} $\mu$.

\section{Statistics}

The number of elementary quanta of action within the volume element $dV$ and the interval of absolute values of the particle momentum from $p$ to $p + dp$ (the number of cells $h^3$ in the phase space of coordinates and momenta for the Planck constant $h$) is given by
\begin{equation}\label{eq:number_of_cells}
	\frac{4\pi p^2 \cdot dp \cdot dV}{h^3}
\end{equation}

According to Fig. \eqref{fig:dV}, the instantaneous energy of a particle at a point within the space element $dV$ is given by
\begin{equation}\label{eq:16}
	\varepsilon(1) = \frac{p^2}{2m} + \varphi(1) \cdot m + \mu,
\end{equation}
where $m$ represents the particle mass, and $\varphi(1)$ denotes the potential of the gravitational field at that particular point. For charged particles in an electric field, the expression becomes
\begin{equation}\label{eq:17}
	\varepsilon(1) = \frac{p^2}{2m} + \varphi(1) \cdot q + \mu,
\end{equation}
where $q$ stands for the particle charge, and $\varphi(1)$ denotes the potential of the electric field at that point.

According to \eqref{eq:homogeneity}, the energy of a single cell $h^3$ occupied by $N$ particles is $N\varepsilon$, representing the total energy of the sub-ensemble. 

For bosons, any number of particles can occupy a single-particle state within a single cell, and the partition function is given by
\begin{equation}\label{eq:18}
	\mathcal{Z} = \sum_{N=0}^{\infty} e^{ -\dfrac{N\varepsilon}{kT} } = \frac{1}{1-e^{ -\dfrac{\varepsilon}{kT} }},
\end{equation}
where $k$ is the Boltzmann constant.

For fermions, a single-particle state can be occupied by one particle or remain empty, and the partition function is expressed as
\begin{equation}\label{eq:19}
	\mathcal{Z} = e^{ -\dfrac{0 \cdot \varepsilon}{kT} } + e^{ -\dfrac{1 \cdot \varepsilon}{kT} }. 
\end{equation}

The quotient of the Boltzmann factor to the normalization constant is given by
\begin{equation}\label{eq:20}
	f = \frac{e^{ -\dfrac{\varepsilon}{kT} }}{\mathcal{Z}} = \frac{1}{e^{ \frac{\varepsilon}{kT} } \pm 1 },
\end{equation}
where '+' is for fermions and '--' is for bosons\footnote{The difference from traditional Fermi-Dirac and Bose-Einstein statistics is due to the absence of particle exchange with the reservoir; the Eqs. \eqref{eq:18}-\eqref{eq:20} are expressions for interacting particles; the notation of chemical potential here is $\mu=\frac{\partial E}{\partial N}$ due to the variational notation $\delta \left\{E -\mu N\right\}=0$, while traditional statistics model the exchange of particles between the system and a reservoir, counting the energy $E_n$ in the system from the value of chemical potential $\mu'$ in the reservoir, so that the Boltzmann factor is $e^{-\frac{E_n-\mu' N}{kT}}$ and the normalization constant is $Z=\sum_N \sum_{n} e^{-\frac{E_n-\mu' N}{kT}}$, that is a grand-canonical ensemble. It is essentially the same as a use of the notation $\delta \left\{E +\mu' N\right\}=0$ and $\mu'=-\frac{\partial E}{\partial N}$, thus in the traditional expressions of Fermi-Dirac and Bose-Einstein statistics, the $\mu$ parameter is subtracted from mechanical energy, while in Eqs. \eqref{eq:16}-\eqref{eq:17} and thus Eqs. \eqref{eq:18}-\eqref{eq:20}, it is with a '+' sign; otherwise, as long as the notations $\delta \left\{E -\mu N\right\}=0$ and $\delta \left\{E +\mu' N\right\}=0$ have no fundamental differences, and the chemical potential $\mu$ or $\mu'$ always appears in differences, Eqs. \eqref{eq:18}-\eqref{eq:20} are equivalent to traditional quantum statistics, yielding the same probabilities of states; here the absolute value of $\mu$ appears in the computation of particle density, and the notation $\delta \left\{E -\mu N\right\}=0$ is used for consistency with other parts. In addition, using a canonical ensemble is easier than using a grand-canonical one. The simplification is only possible due to the variational argument \eqref{eq:homogeneity}; }.

The partition functions allow to calculate the expected number of particles in a single-particle state in cell $h^3$ as follows:
\begin{equation}\label{eq:21}
	\big\langle  N \big\rangle = \sum_{N} N P_N=\frac{1}{\mathcal{Z}} \sum_{N} N \, e^{ -\dfrac{ \varepsilon N}{kT} } = -kT \frac{1}{\mathcal{Z}} \dfrac{\partial \mathcal{Z}}{\partial \mu} = f,
\end{equation}
and the expected value of energy is given by
\begin{equation}\label{eq:22}
	\big\langle  E \big\rangle = \sum_{N} E_N P_N=\frac{1}{\mathcal{Z}} \sum_{N} (\varepsilon N) \, e^{ -\dfrac{ \varepsilon N}{kT} } = -\frac{1}{\mathcal{Z}} \dfrac{\partial \mathcal{Z}}{\partial \beta} = f \varepsilon,
\end{equation}
where $\beta = \frac{1}{kT}$.\\

The Gibbs entropy is given by
\begin{multline}\label{eq:23}
	S = -k\sum_{N} P_N \ln P_N= - k \sum_{N} \frac{e^{ -\dfrac{ \varepsilon N}{kT} }}{\mathcal{Z}}  \ln \frac{e^{ -\dfrac{ \varepsilon N}{kT} }}{\mathcal{Z}}=\\ k \sum_{N} \frac{e^{ -\dfrac{ \varepsilon N}{kT} }}{\mathcal{Z}} \Bigg( \dfrac{ \varepsilon N}{kT} + \ln \mathcal{Z} \Bigg)	= \frac{\big\langle  E \big\rangle}{T} + k \ln \mathcal{Z}.
\end{multline}
The entropy expression \eqref{eq:23} in the literature is called entropy for the canonical ensemble, regarded in terms of the canonical partition function \eqref{eq:18}. On the other hand, Eqs. \eqref{eq:19}-\eqref{eq:20} and \eqref{eq:21}-\eqref{eq:22} are commonly considered properties of the grand-canonical ensemble, which involves the exchange of particles with a reservoir. However, the definition of chemical potentials \eqref{eq:chem_potentials} with the property \eqref{eq:homogeneity} makes the exchange of particles with a reservoir redundant and simplifies the statistical expressions. Now, Eqs. \eqref{eq:19}-\eqref{eq:20} and \eqref{eq:21}-\eqref{eq:22} are properties of the canonical ensemble of interacting particles, with all interactions accounted for in the values of equalized chemical potentials.

\section{The temperature of elementary cell $h^3$}
The entropy expression \eqref{eq:23} implies that the single-particle state within a single cell $h^3$ has a temperature $T$ determined by the heat reservoir. Nevertheless, it is noteworthy that the above definition of chemical potentials and the accompanying variational arguments make heat exchange with the reservoir redundant too. The indeterminacy and statistical stationarity may be properties of the system itself.

Explicit accounts for the temperature of $h^3$ due to occupation by a particle or interactions of occupying particles with fields can be described as follows. The Boltzmann entropy $S = k \ln \Omega$ is expressed through the ensemble $\Omega$. From Eq. \eqref{eq:23}, the ensemble of single-particle states in a single cell $h^3$ reads
\begin{equation}\label{eq:entropy_ensemble}
	\Omega = \mathcal{Z} \, e^{\frac{\langle E \rangle}{kT}}.
\end{equation}
If a change of energy $\langle E \rangle = f \epsilon$ by $d\langle E \rangle$ is caused by the change of the population of the elementary cell $h^3$ or by a change in the potential $\varphi$ in Eqs. \eqref{eq:16}-\eqref{eq:17}, then there is a corresponding change of entropy $dS = k\, d\big(\ln\Omega\big)$, and $\frac{1}{T} = \frac{dS}{d\langle E \rangle}$.

\section{Sources of force fields}
Particles of mass $m$ or charge $q$ are sources of gravitational and electric fields, which depend on the number density of particles. Each cell $h^3$ may contain some number of particles with energy $\varepsilon$. This is recognized by the density of states $g$, which relates to the nature of particles, the corresponding possible number of particles within a cell, and other factors. The phase-space-based number density of particles then reads
\begin{equation}\label{eq:25}
	\rho(1) = \frac{4\pi}{h^3} \int_{0}^{+\infty} \frac{ g \cdot p^2 }{e^{ \frac{\varepsilon}{kT} } \pm 1} \cdot dp,
\end{equation}
where Eqs. \eqref{eq:number_of_cells} and \eqref{eq:20} were used\footnote{the number of cells \eqref{eq:number_of_cells} for particular values $x,\,y,\,z$ and $p$ is multiplied by the probability $f$ to find a particle in such cell, and summed for all possible momenta, $g$ accounts for the possibility of several particles in a single cell;}. The number of particles $dN$ within the space element $dV$ is undefined until the observation procedure for the number of particles is conducted, while the density of the particles is predefined\footnote{this density describes the system of particles <<in small>> as well as it can not be directly normed on the total number of particles while accounting for the heat behavior of particles within the volume element $dV$ with a local field by Eqs. \eqref{eq:16}-\eqref{eq:17}, nevertheless it can be normed on the total number of particles in the system by the Eq. \eqref{eq:normalization}, which yields the value of chemical potential $\mu$;\label{footnote_37}} by the statistical law \eqref{eq:25}.

The source function \eqref{eq:25} is investigated for gravitation (Eq. \eqref{eq:16}) and for electric field (Eq. \eqref{eq:17}). Fermions are usually associated with matter, and the '+' sign in Eq. \eqref{eq:25} is of most practical interest here. For fermions with a spin $\pm \frac{1}{2}$, the value $g=2$ is taken below according to the Pauli exclusion principle.

For $T=0\, K$ the denominator of the Eq. \eqref{eq:25} takes value $\pm1$ for $\varepsilon<0$ and infinite value if $\varepsilon > 0$. The equation $\varepsilon = 0$ defines a Fermi sphere in the phase space of coordinates and momenta, where particles attain the highest possible momentum at zero temperature, $p_{max}(1)=\sqrt{2m(-\varphi(1)\cdot m - \mu)}$ for gravitational field\footnote{$\frac{p^2}{2m}+\varphi(1)\cdot m+\mu \le 0 \Rightarrow p_{max}(1)=\sqrt{2m(-\varphi(1)\cdot m - \mu)}$, for gravitation $\varphi(1) \le 0$ and the whole expression under square root is non-negative;}, and the integral \eqref{eq:25} turns for a finite range of momenta of particles at gravitational field to\footnote{$\rho(1) = \frac{4 \pi }{h^3} \int_{0}^{p_{max}(1)} g \cdot p^2\, dp = \frac{16 \pi \sqrt{2} m^{\frac{3}{2}}}{3h^3}\cdot\big(-\varphi\cdot m - \mu\big)^\frac{3}{2}$ for $g=2$;} 
\begin{equation}\label{eq:26}
	\rho(1) = \frac{16 \pi \sqrt{2} m^{\frac{3}{2}}}{3h^3}\cdot\big(-\varphi(1)\cdot m - \mu\big)^\frac{3}{2}, \qquad T= 0\,K.
\end{equation}

For electrons at the electric field $p_{max}(1)=\sqrt{2(\varphi(1) - \mu)}$ in atomic units and\footnote{in atomic units the Planck constant is $h = 2 \pi$, the elementary charge is $e = 1$, the charge of an electron is $q_e = -1$, the mass of an electron is $m_e = 1$, and the Coulomb constant $\frac{1}{4 \pi \epsilon_0} = 1$. The Eq. \eqref{eq:number_of_cells} in atomic units is $\frac{p^2 \cdot dp \cdot dV}{2 \pi^2}$, the \eqref{eq:17} reads $\varepsilon(r) = \frac{p^2}{2} - \varphi(r) + \mu$, $\frac{p^2}{2} - \varphi(r) + \mu \le 0 \Rightarrow p_{max}(1)=\sqrt{2(\varphi(1) - \mu)}$, the integral  \eqref{eq:25} in atomic units is $\rho(r) = \frac{1}{2\pi^2} \int_{0}^{+\infty} \frac{ g \cdot p^2 }{e^{ \frac{\varepsilon}{kT} } + 1} \cdot dp$, and for $T=0\, K$ for finite range of momenta of electrons $\rho(1) = \frac{1 }{\pi^2} \int_{0}^{p_{max}(1)} p^2\, dp = \frac{2 \sqrt{2}}{3 \pi^2}\cdot\big(\varphi(1) - \mu\big)^\frac{3}{2}$ for $g=2$, for positive electric field $\varphi(1) \geq 0$ and the whole expression under square root is non-negative;}
\begin{equation}\label{eq:27}
	\rho(1) = \frac{2 \sqrt{2}}{3 \pi^2}\cdot\big(\varphi(1) - \mu\big)^\frac{3}{2}, \qquad T= 0\,K,
\end{equation}

\section{The gravitational field of a hot source}

The mass of the star consists almost entirely of protons and neutrons. A Poisson equation $\Delta \varphi = 4\pi G \cdot m\rho$ for a spherically symmetric gravitational field reads\footnote{here in the right part of the Poisson equation, the phase-space-based mass density $m \cdot \rho$ is produced from the phase-space-based number density $\rho$;}
\begin{equation}\label{eq:28}
	\frac{1}{r}\frac{d^2}{dr^2}\Big(r\,\varphi(r)\Big) = 4 \pi G \cdot m \cdot  \frac{8\pi}{h^3} \int_{0}^{+\infty} \frac{  p^2 }{e^{ \frac{\varepsilon}{kT} } + 1} \cdot dp,\quad  0 < r < +\infty,
\end{equation}
where $G$ is the universal gravitational constant. Two boundary conditions are $\varphi(+\infty) = -\frac{\mu}{m}$ and $\varphi(r) \sim -G\frac{M}{r}$ when $r \rightarrow 0$, where $M$ is the mass of the source\footnote{the reference point $\varphi(+\infty) = -\frac{\mu}{m}$ of gravitational potential reflects the necessary consistency of references for a force field potential and chemical potential as already mentioned in a footnote \eqref{footnote_34}, the density of particles Eq. \eqref{eq:26} vanishes at infinity for the $\varphi(+\infty) = -\frac{\mu}{m}$, the same can be shown for the right side of the Eq. \eqref{eq:28} for $T > 0\,K$;}.

In astronomical units (AU), the universal gravitational constant is $G = 4\pi^2$, where $\left[m\right] = 1 M_{\odot} = 1.98892\cdot10^{30}\, kg$ is the mass of the Sun, $\left[L\right] = 1 \, AU$ is the distance between the Sun and the Earth, fixed as 149597870700 meters, $\left[t\right] = 1\, year$ fixed as $86400 \cdot 365.25$ seconds. The speed of light in astronomical units is $c = 6.324107708 \cdot 10^4 \, \frac{AU}{year}$. The astronomical unit of energy is $\left[E\right] = 1\,M_{\odot} \cdot c^2 = 1 \, (AUE) = 3.9994\cdot10^9 \, \frac{M_{\odot} \cdot AU^2}{year^2}$. Boltzmann constant is $k = 3.089037\cdot10^{-61}\,(AUE)\cdot K^{-1}$. And Planck's constant is $h = 4.69777\cdot10^{-79} \, (AUE) \cdot year$. The mass of a neutron in astronomical units is $m_n = 8.421291445\cdot10^{-58}\, M_{\odot}$.

Variable substitution $\mathcal{X}(r) = \frac{r}{G}\big(\varphi(r) + \frac{\mu}{m}\big)$ turns the Eq. \eqref{eq:28} into a common two-point boundary problem, where $\mathcal{X}(0) = -M$ and $\mathcal{X}(+\infty) = 0$. To work with the integral on the right side the Fermi-Dirac special function $F_{\frac{1}{2}}$ is introduced
\begin{equation}\label{eq:F05}
	F_{\frac{1}{2}}(\eta) = \int_{0}^{+\infty} \frac{ \sqrt{x} }{e^{ x - \eta } + 1} \, dx.
\end{equation}

The Eq.\eqref{eq:28} reads\footnote{on left side $\frac{1}{r}\frac{d^2}{dr^2}\Big(r\,\varphi(r)\Big) = \frac{G}{r}\frac{d^2 \mathcal{X}(r)}{dr^2}$ is used; on right side for $T > 0\,K$, $\int_{0}^{+\infty} \frac{  p^2 }{e^{ \frac{1}{kT}\left(\frac{p^2}{2m}+\varphi m +\mu\right) } + 1} \cdot dp = \sqrt{2}(kT)^\frac{3}{2}\cdot m^\frac{3}{2} \int_{0}^{+\infty} \frac{  \frac{p}{\sqrt{2kTm}} }{e^{ \left(\frac{p^2}{2kTm}- \frac{1}{kT}\left( -\varphi m -\mu\right)\right) } + 1} \cdot d\big(\frac{p^2}{2kTm}\big)=\sqrt{2}(kT)^\frac{3}{2}\cdot m^\frac{3}{2} F_\frac{1}{2}\big(-\frac{\varphi m + \mu}{kT}\big)=\sqrt{2}(kT)^\frac{3}{2}\cdot m^\frac{3}{2}F_{\frac{1}{2}}\Big(- \frac{Gm\mathcal{X}(r)}{r\,kT}\Big)$; on right side for $T = 0\,K$, $\rho(r) = \frac{16 \pi \sqrt{2} m^{\frac{3}{2}}}{3h^3}\cdot\big(-\varphi(r)\cdot m - \mu\big)^\frac{3}{2}=\frac{16 \pi \sqrt{2} m^{\frac{3}{2}}}{3h^3}\cdot\big(-\mathcal{X}(r) \frac{G}{r} m \big)^\frac{3}{2}=16 \pi \sqrt{2} \frac{m^3}{h^3} \frac{G^\frac{3}{2}}{r^\frac{3}{2}}(-\mathcal{X}(r))^\frac{3}{2}$;\label{footnote_43}}
\begin{equation}\label{eq:30}
	\dfrac{d^2\mathcal{X}(r)}{dr^2} =
	\begin{cases}
	 4 \pi \cdot \frac{m^\frac{5}{2}}{h^3} \cdot 8 \pi \sqrt{2} \cdot (kT)^\frac{3}{2} \cdot   rF_{\frac{1}{2}}\Big(- \frac{Gm\mathcal{X}(r)}{r\,kT}\Big), \quad T > 0\,K,		
	 \\
	 4 \pi \cdot \frac{m^4}{h^3} \cdot G^\frac{3}{2} \cdot \frac{16 \pi \sqrt{2}}{3}\cdot  \frac{(-\mathcal{X}(r))^\frac{3}{2}}{\sqrt{r}},\quad  T = 0\,K.
	\end{cases}
\end{equation}

The ratio $m ^\frac{5}{2} / h^3$ is extraordinarily big for any known particles in virtue of values of physical constants in astronomical units\footnote{for instance, for neutron mass $\frac{m^\frac{5}{2}}{h^3}=1.98\cdot 10^{92}$, the right side deviates from the Dirac function due to the quantization of the phase space;}. No matter what particles the source consists of, the Eq. \eqref{eq:28} at any physically possible temperature $T$ is very close to the  $\Delta \varphi(r) = 4 \pi G \cdot \delta(r)$ for Dirac delta function $\delta(r)$, and hence any cold or hot dense source of gravitation is described by Newton's law of universal gravitation
\begin{equation}\label{eq:31}
	\varphi(r) = -G \cdot \frac{M}{r} -  \frac{\mu}{m},
\end{equation}
where the origin of potential $-\mu / m$ was introduced according to remarks for the consistency of force field potential and chemical potential, to provide vanishing of the number density of particles at infinitely distanced points from any sources of forces.

Deviations from the law \eqref{eq:31} appear when the point source is located in ambient gas of the temperature $T$ and heat processes in the gas affect the potential of the gravitation field.

To account for the lack of particles for continuous filling of cells $h^3$ of phase-space of coordinates and momenta, we take a properly small density $gK$ in place of $g$ in the Eq. \eqref{eq:25}, where $K \ll 1$. The gas is homogeneous in the phase space of coordinates and momenta with a small density of states $gK$ reflected by the constant $c=\frac{4\pi gK m^\frac{3}{2} \sqrt{2}}{h^3}$ in the number density of particles around a massive source, and has thermal-statistical properties for particles of the source\footnote{the expression for constant $c$ is motivated by $K\rho(1)=\frac{4\pi K}{h^3}\int_{0}^{+\infty} \frac{ g p^2 }{e^{ \frac{\varepsilon}{kT} } + 1} \cdot dp = \big( \frac{4\pi gK m^\frac{3}{2} \sqrt{2}}{h^3}  \big) (kT)^\frac{3}{2} F_\frac{1}{2}\big(-\frac{\varphi(1) \cdot m + \mu}{kT}\big)$; as mentioned in a footnote \ref{footnote_23}, the phase-space-based number density of particles can not be normed to any given number of particles, for instance, if a cell $h^3$ can be occupied by two particles with spins $\pm \frac{1}{2}$, then $g=2$, and for any normalization factor $K$ greater than one $gK > 2$ and the cell might be occupied by more than two particles in contradiction with the Pauli exclusion principle, but the phase-space-based density of particles can be scaled down for $K \le 1$ without contradiction with the Pauli exclusion principle, that yields the model of gas here; the factor $\frac{2}{3}$ for $T=0\,K$ appears after the substitution $gK=\frac{c \cdot h^3}{4 \pi m^\frac{3}{2} \sqrt{2}}$ to $\rho(1) = \frac{8 \pi \sqrt{2} m^{\frac{3}{2}}\,gK}{3h^3}\cdot\big(-\varphi(1)\cdot m - \mu\big)^\frac{3}{2}$;}.

\begin{equation}\label{eq:32}
	\rho(r) = 
	\begin{cases}
			c \cdot (kT)^\frac{3}{2} \, F_{\frac{1}{2}}\Big(-\frac{\varphi(r) \cdot m + \mu}{kT}\Big), \quad T > 0\,K,
		\\
		\frac{2}{3} c \left(-\varphi(r) \cdot m - \mu\right)^\frac{3}{2},\quad  T = 0\,K.
	\end{cases}
\end{equation} 

The Poisson equation for the gravitational field of a point source in rarefied ambient gas reads
\begin{equation}\label{eq:33}
	\dfrac{d^2\mathcal{X}(r)}{dr^2} = 
	\begin{cases}
4 \pi \cdot c \cdot m \cdot (kT)^\frac{3}{2} \, rF_{\frac{1}{2}}\Big(-\frac{G m \mathcal{X}(r)}{r\,kT}\Big), \quad T > 0\,K,
		\\
4 \pi \cdot \frac{2}{3}\cdot c \cdot  G^\frac{3}{2}\, m^\frac{5}{2}\, \frac{1}{\sqrt{r}}
\left(-\mathcal{X}(r)\right)^\frac{3}{2},\quad  T = 0\,K
	\end{cases}	
\end{equation}
where $\mathcal{X}(0) = -M_{\odot}$ for the source with mass of the Sun and $\mathcal{X}(+\infty) = 0$. The graph of the $F_{\frac{1}{2}}$ function as well as a method for solving such an equation by the Euler broken line method are given in the appendix.

A solution of the Poisson equation $\varphi(r) = G \frac{\mathcal{X}(r)}{r} - \frac{\mu}{m}$ at radius $r$ represents the potential corresponding to mass contained within a sphere of radius $r$ that provides the tension of gravitational field $-\frac{d \varphi(r)}{dr} = -G \frac{M_{\odot} + M_r}{r^2}$, where $M_r = m \int_{0}^{r} \rho( r ) \cdot 4 \pi r^2\cdot dr$ is the mass of gas in the sphere of radius $r$ around the point source. Since this gas density cannot be normed to any given number of particles, the mass $M_r$ is the apparent mass of the gas in the sphere of radius $r$, it depends on the gas temperature for a fixed parameter $c$. Integral 
\begin{equation}\label{eq:apparent_mass}
	M_\infty = m \cdot 4 \pi \int_{0}^{\infty} \rho( r ) \cdot r^2\cdot dr
\end{equation}
represents total \textit{apparent} mass of gas, while the normalization of the density $m\rho$ for a given real mass of particles $M_{gas}$ in the gas has to be done by the Eq. \eqref{eq:normalization} in terms of assigning a proper value to the chemical potential $\mu$ and correspondingly to the reference point $-\frac{\mu}{m}$ of the gravitational potential $\varphi(r)$. Suppose the normalization equation as $\mu N = \int_{0}^{\infty} m\rho(r) \varphi(r) \cdot 4 \pi r^2 \cdot dr$ and real number of particles in the gas $N = \frac{M_{gas}}{m}$, it follows\footnote{here $\mu$ is expressed from $\mu \frac{M_{gas}}{m} =\int_{0}^{\infty} m \rho(r) \big(G \frac{\mathcal{X}(r)}{r} - \frac{\mu}{m}\big) \cdot 4\pi r^2 \cdot dr$;}
 \begin{equation}\label{eq:35}
	\mu =
	\begin{cases}
			 \frac{m^2\cdot c \cdot 4\pi G \cdot (kT)^\frac{3}{2}\,  \int_{0}^{+\infty} F_{\frac{1}{2}}\big(-\frac{Gm \mathcal{X}(r)}{r\,kT}\big) \mathcal{X}(r)\, rdr}{M_{gas} + m \cdot c \cdot 4\pi \cdot (kT)^\frac{3}{2}\, \int_{0}^{+\infty} F_{\frac{1}{2}}\big(-\frac{Gm\mathcal{X}(r)}{r\,kT}\big) \, r^2dr}, \quad T > 0\,K,
		\\
		-\frac{\frac{2}{3}\cdot c \cdot G^\frac{5}{2}m^\frac{7}{2} 4 \pi \int_{0}^{\infty}\frac{\left(\mathcal{X}(r)\right)^\frac{5}{2}}{\sqrt{r}}dr}{M_{gas}+\frac{2}{3} \cdot c \cdot G^\frac{3}{2}m^\frac{5}{2} 4 \pi \int_{0}^{\infty} \left(\mathcal{X}(r)\right)^\frac{3}{2} \sqrt{r} dr},\quad  T = 0\,K.
	\end{cases}
\end{equation}

The field $\varphi(r) = G\frac{\mathcal{X}(r)}{r} - \frac{\mu}{m}$ possesses a temperature $T$, and the energy of mass density of gas cannot be calculated by mechanical means\footnote{it possesses a temperature $T$ by Eq. \eqref{eq:entropy_ensemble};}. The expression for the energy of mass density in the \textit{outer}\footnote{there was no expression for the derivative $\frac{\delta F}{\delta \rho}$ in the normalization Eq. \eqref{eq:normalization} and the local homogeneous integral functional on its right side was supposed $\int m\rho \varphi\, dV$, as energy in \textit{outer} potential -- without a factor $\frac{1}{2}$ for the potential energy of mass density in a gravitational field; obviously, the choice of a factor for the integral $\int m\rho \varphi\, dV$ affects only the normalization of the computed energy, and not on the field $-\nabla \varphi$; the \textit{absolute} value of the chemical potential, based on the choice $\mu N=\int m\rho \varphi\, dV$ is consistent with classical Thomas-Fermi theory and yields energies for many-electron systems matched with Hartree-Fock theory;\label{footnote_48}} potential field $\int m\rho \varphi\, dV$ is undefined due to the arbitrariness of the origin point of the potential $\varphi$ until the value of $\mu$ is assigned. While mechanical calculations might suppose $\mu = 0$, for particles in heat equilibrium, this assumption would be equivalent to supposing zero internal energy for both the ambient gas and the star inside it, as dictated by Eq. \eqref{eq:homogeneity}. Therefore, for an accurate calculation of energy, the value of the chemical potential $\mu$ has to be determined in terms of heat equilibrium. This equation \eqref{eq:35} provides a means to calculate the energy of a density of particles in gas with temperature $T$ as $U=\mu N$.

At any temperature, the chemical potential Eq. \eqref{eq:35} for gravitation is extraordinarily small in virtue of the value of Bolzman constant $k$ in astronomical units, and the mass $m$ of a particle. Newton's law of universal gravitation \eqref{eq:31} holds in the vicinity of any point source, regardless of the density of ambient gas, as long as the principal part of the potential is $\varphi(r) \sim -G\frac{M}{r}$. When the law is observed, the tiny value of $-\frac{\mu}{m}$ is not recognizable. For the vanishing density of ambient gas $c \rightarrow 0$, the origin of the gravitational potential $-\frac{\mu}{m}$ vanishes in the law of gravitation \eqref{eq:31}.

The chemical potential is defined and the heat equilibrium of the system of particles can be considered.  To solve the equations \eqref{eq:33} the coefficient value $c$ is required. Its value for $K \ll 1$ is not crucial for the qualitative picture of heat phenomena on gravitation\footnote{for neutron mass $m$ the $c=\frac{4\pi gK m^\frac{3}{2} \sqrt{2}}{h^3} = K \cdot 8.38 \cdot 10^{150}$; if to suppose $K = 1$ for the matter of neutron star, where mass density is $m\rho = 10^{15}\, \frac{g}{cm^3}$ then $K \sim \frac{m \rho_{gas}}{m \rho_{neutron star}}$, for hydrogen at normal conditions it might be assumed $c = 8.38 \cdot 10^{133}$, for extremely rarefied gas between stars  with $10^6$ particles per $1\,cm^3$ it would be $c = 8.38 \cdot 10^{115}$, and for the most rarefied gas in the Universe, supposed for a pressure $10^{-17}\,Pa$ it would be $c = 8.38 \cdot 10^{106}$;of course, these values are arbitrary, since the coefficient $c$ is intended to characterize the rarefaction in the phase space of coordinates and momenta, and not just in spatial variables; \label{footmark_49}}. 

The field determined by the Poisson equation \eqref{eq:33} for the distribution of particles in phase space depends on temperature, just as for thermal effects in an ideal gas the field potential depends on the temperature in the system \eqref{eq:thermal_effects}. The mass illusion appears due to the fact, that the integral of the mass density of gas on the right side of the Poisson equation is not equal to the real mass, for which the chemical potential \eqref{eq:35} was defined. The apparent mass of gas \eqref{eq:apparent_mass} is small for low temperature\footnote{it shows, that assuming some relatively large density coefficient $c$ we do not suppose dense matter around the point source, but suppose rarefied gas;} but appears to be astronomically large for high temperature of a gas, Tab. \eqref{tab:table1}. The apparent mass $M_\infty$ doesn't affect bodies in the field near the point source like a mass of a hollow ball doesn't affect a body inside of it. The apparent mass relates to the heat in the gas\footnote{in particular, it doesn't have the Einstein energy equivalent $M_\infty C^2$ for the speed of light $C$, as obviously from values of energy $U$ in Tab. \eqref{tab:table1}, and the denominator of the expression \eqref{eq:35} for the chemical potential is a sum $M_{gas} + M_\infty$ of real mass of the gas and its apparent mass; } and affects the gravitational potential $\varphi(r)$.

 \renewcommand{\tablename}{Tab.}
\begin{table}[!h]
	\caption{\label{tab:table1} Temperature dependence of the gravitational field  of point mass $1\,M_{\odot}$ in ambient gas of mass $1\,M_{\odot}$.}
	\begin{center}
		\begin{tabular}{|c||c|c|c|c|}
			\hline
			$\lg c$ & $T$,\,K & $\mu,\,AU$ & $U=\mu N,\,AU$ & $M_\infty,\,AU$ \\
			\hline
			\multirow{3}{2em}{130} & 0 & $-2.971\cdot10^{-64}$ & $-3.528\cdot10^{-7}$ & $8.942\cdot10^{-7}$  \\
			& $10^5$ & $-5.203\cdot10^{-61}$ & $-6.178\cdot10^{-4}$ & 0.016 \\
			 & $10^6$ & $-1.306\cdot10^{-59}$ & $-0.0155$ & 0.513 \\
			\hline	
			\multirow{3}{2em}{131} & 0 & $-2.964\cdot10^{-63}$ & $-3.520\cdot10^{-6}$ & $8.902\cdot10^{-6}$  \\
            & $10^5$ & $-4.816\cdot10^{-60}$ & $-5.719\cdot10^{-3}$ & 0.162 \\
            & $10^6$ & $-7.472\cdot10^{-59}$ & $-8.873\cdot10^{-2}$ & 5.133 \\
            \hline
			\multirow{3}{2em}{132} & 0  & $-2.963\cdot10^{-62}$ & $-3.519\cdot10^{-5}$ & $8.899\cdot10^{-5}$ \\
			& $10^5$  & $-3.161\cdot10^{-59}$ & $-3.754\cdot10^{-2}$ & 1.623 \\
			 & $10^6$  & $-5.596\cdot10^{-58}$ & $-0.665$ & 51.334 \\
			\hline
			\multirow{3}{2em}{133} & 0  & $-2.968\cdot10^{-61}$ & $-3.525\cdot10^{-4}$ & $8.942\cdot10^{-4}$ \\
             & $10^5$  & $-2.007\cdot10^{-58}$ & $-0.238$ & 16.233 \\
             & $10^6$  & $-5.912\cdot10^{-58}$ & $-0.702$ & 513.34 \\
            \hline			
		\end{tabular}
	\end{center}
\end{table}

\begin{figure}
	\begin{center}
		\begin{tikzpicture}
			\begin{axis}[
				table/col sep = semicolon,
				width = 400pt, height = 250pt,
				xmin = 0,
				xmax = 10,
				ymax = -0.8,
				ymin = -1.4, 
				xlabel = {\normalsize $r, \; AU$}, 
				ylabel = {\normalsize $\mathcal{X}(r)$}
				]
				\addplot[mark options={scale = 1.5}] table [x={x}, y={y}] {c0_Sun_0_R500.csv} node[pos=0.92,yshift=-0.2cm]{$c = 0$};
				\addplot[mark options={scale = 1.5}] table [x={x}, y={y}] {c132_Sun_0_R500.csv};
				\addplot[mark options={scale = 1.5}] table [x={x}, y={y}] {c132_Sun_10_5_R500.csv}node[pos=0.8,yshift=0.3cm,rotate=2]{$T=0 \div 10^5\,K$};				
				\addplot[mark options={scale = 1.5}] table [x={x}, y={y}] {c132_Sun_10_6_R500.csv}node[pos=0.8,yshift=0.3cm,rotate=-22]{$T=10^6\,K$};
			\end{axis}
		\end{tikzpicture}
		\caption{\label{fig:2} The gravitational field of point mass $1\,M_{\odot}$ in ambient gas for $\lg c = 132$.}	
	\end{center}
\end{figure}
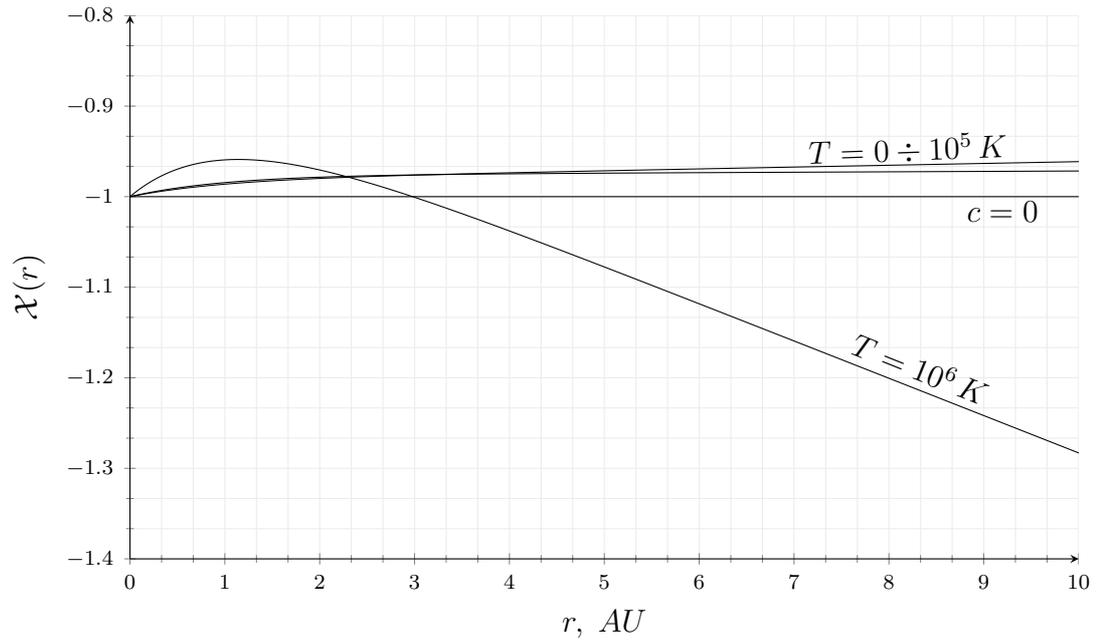

\begin{figure}
	\begin{center}
		\begin{tikzpicture}
			\begin{axis}[
				table/col sep = semicolon,
				width = 400pt, height = 250pt,
				xmin = 0,
				xmax = 10,
				ymax = 0.0,
				ymin = -1.2, 
				xlabel = {\normalsize $r, \; AU$}, 
				ylabel = {\normalsize $\mathcal{X}(r)$}
				]
				\addplot[mark options={scale = 1.5}] table [x={x}, y={y}] {c0_Sun_0_R500.csv} node[pos=0.5,yshift=-0.2cm]{$c = 0$};
				\addplot[mark options={scale = 1.5}] table [x={x}, y={y}] {c133_Sun_0_R500.csv}node[pos=0.9,yshift=0.3cm]{$T=0\,K$};				
				\addplot[mark options={scale = 1.5}] table [x={x}, y={y}] {c133_Sun_10_5_R500.csv}node[pos=0.3,yshift=0.3cm]{$T=10^5\,K$};				
				\addplot[mark options={scale = 1.5}] table [x={x}, y={y}] {c133_Sun_10_6_R500.csv}node[pos=0.8,yshift=0.3cm,rotate=-5]{$T=10^6\,K$};
			\end{axis}
		\end{tikzpicture}
		\caption{\label{fig:3} The gravitational field of point mass $1\,M_{\odot}$ in ambient gas for $\lg c = 133$.}	
	\end{center}
\end{figure}
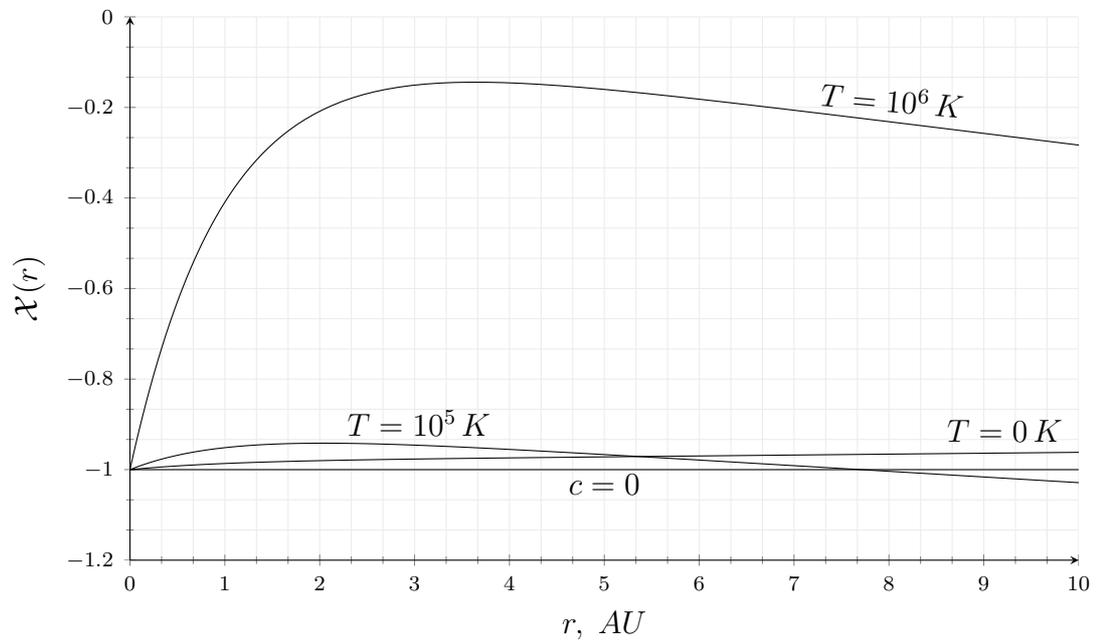

Gravitational fields of point mass $1\,M_{\odot}$ for different states of ambient gas are shown in Figs. \eqref{fig:2},\eqref{fig:3}. The horizontal curve represents Newton's law of universal gravitation \eqref{eq:31} for $c = 0$, where the Eq. \eqref{eq:33} reduces to the Laplace equation. For a non-zero density of gas at zero temperature the solution is close to Newton's law of universal gravitation, while for a large positive temperature, it deviates. For smaller density coefficient $\lg c < 130$ the qualitative picture is similar to the Fig. \eqref{fig:2}, while for bigger $\lg c \gg 130$ for a dense ambient gas, all solutions for all temperatures systematically close to fields predicted by Thomas-Fermi theory \cite{Thomas} of the atom -- for monotonically with $r$ and fast approaching zero density of particles in a gas. 

\section{The electric field of a hot source}

An atomic core is a point source of a positive electric field, surrounded by a gas of electrons. A common abstraction for the electron number density energy functional is $E[\, N,\,\rho\,] = (-1)\cdot\int v \rho \, dV + F[\, N,\,\rho\,]$, where $v$ is an outer potential\footnote{the charge density $(-1)\cdot\rho$ in atomic units is produced here from the number density $\rho$;} due to nuclei, and $F$ represents the unknown but well-defined kinetic and electron-electron interaction energy for the number density $\rho$.
 
It is immediately known that when the stationarity necessary conditions are met, the functional $E[\, N,\,\rho\,]$ turns into a homogeneous local integral functional of degree one, given by $E[\, N,\,\rho\,] = \int \frac{\delta E}{\delta \rho} \rho \, dV = -\int \varphi \rho \, dV$, where $\frac{\delta E}{\delta \rho}$ is denoted as $-\varphi$ because electrons have negative charge, and the integration is over the space occupied by the density. The derivative $\frac{\delta E}{\delta \rho}$ is not given and we apply the phase-space-based number density \eqref{eq:25}. For electrons in atomic units\footnote{Boltzman constant in atomic units is $k = 3.166811563\cdot 10^{-6} \, a.u. / K$;} it is
\begin{equation}\label{eq:36}
	\rho(1) = \frac{1}{2\pi^2} \int_{0}^{+\infty} \frac{  g\cdot p^2 }{e^{ \frac{1}{kT}\left(\frac{p^2}{2}-\varphi(1) +\mu\right) } + 1} \cdot dp,
\end{equation}
where $g=2$ for many-electron systems, and $g=1$ for a one electron\footnote{we apply a statistical expression for the number density after we investigated the state of a single cell $h^3$ with Eq. \eqref{eq:18} as long as the homogeneity \eqref{eq:homogeneity} no more relies on a statistically large number of particles, but if there is a single electron, no one cell can be occupied by two, and thus $g=1$ for a one-electron system;}. The density is summed from phase-space cells $h^3$ each of which contains an undefined number of electrons not exceeding $g$ and has a temperature $T$ by Eq. \eqref{eq:entropy_ensemble}. The potential $\varphi$ then is found from electrostatic Poisson equation $\triangle \varphi = -4\pi \cdot (-1)\cdot\rho$, while the chemical potential of cells $h^3$ is defined by \eqref{eq:chem_potentials}, $\mu = \frac{\partial E}{\partial N}= \frac{\delta E}{\delta \rho}$. To compute the energy $E[\, N,\,\rho\,]$ we have to norm it to the value of the chemical potential defined. Up to a constant multiplier as mentioned in the footnote \eqref{footnote_48} the normalization Eq. \eqref{eq:normalization} is $\mu N = -\int \varphi \rho \, dV$.

An electric field in an atom is spherically symmetric. For the consistency of electric field potential $\varphi(r)$ and chemical potential $\mu$ we supposed\footnote{it can be shown the density \eqref{eq:36} for positive temperature $\rho(+\infty)=0$ if $\varphi(+\infty) = \mu$, and $\rho(+\infty)>0$ if $\varphi(+\infty) \ne \mu$;} in the Eqs. \eqref{eq:17} and \eqref{eq:27} that $\varphi(+\infty) = \mu$. The principal part of the potential $\varphi(r)$ is $\varphi(r) \sim \frac{Z}{r}$ when $r \rightarrow 0$, where $Z$ stands for atomic core positive charge attracting electrons. A variable substitution $\mathcal{X}(r) = r\big(\varphi(r) - \mu\big)$ turns the asymptotic boundary condition near $r=0$ to the numeric $\mathcal{X}(0)=Z$, while $\mathcal{X}(+\infty)=0$. After the transformations detailed in the footnote \eqref{footnote_43} the electrostatic Poisson equation reads
\begin{equation}\label{eq:37}
	\dfrac{d^2\mathcal{X}(r)}{dr^2} =
	\begin{cases}
		\frac{g \cdot 2  \sqrt{2}}{\pi} \cdot (kT)^\frac{3}{2} \cdot   rF_{\frac{1}{2}}\Big(\frac{\mathcal{X}(r)}{r\,kT}\Big), \quad T > 0\,K,
		\\
		\frac{g \cdot 4 \sqrt{2}}{3 \pi} \cdot  \frac{\mathcal{X}^\frac{3}{2}(r)}{\sqrt{r}},\quad  T = 0\,K,
	\end{cases}
\end{equation}
with boundary conditions $\mathcal{X}(0)=Z$, $\mathcal{X}(+\infty)=0$, and the electrostatic potential in atom at low temperatures, or generally, the electrostatic potential around atomic core at any temperature is  $\varphi(r) =\frac{\mathcal{X}(r)}{r} + \mu$.

Chemical potential for electrons in the field of atomic core  for $N=Z$ is
\begin{equation}\label{eq:38}
	\mu =
	\begin{cases}
	 - \frac{ \frac{2\sqrt{2}\,g}{ \pi}  \cdot (kT)^\frac{3}{2}\,  \int_{0}^{+\infty} F_{\frac{1}{2}}\Big(\frac{\mathcal{X}(r)}{r\,kT}\Big) \mathcal{X}(r)\, rdr}{Z + \frac{2\sqrt{2}\,g}{\pi} \cdot (kT)^\frac{3}{2}\, \int_{0}^{+\infty} F_{\frac{1}{2}}\Big(\frac{\mathcal{X}(r)}{r\,kT}\Big) \, r^2dr}, \quad T > 0\,K,
		\\
	 -\frac{\frac{4\sqrt{2}\,g }{3\pi}\int_{0}^{+\infty}\frac{\mathcal{X}^\frac{5}{2}(r)}{\sqrt{r}}\,dr}{Z+\frac{4\sqrt{2}\,g }{3\pi}\int_{0}^{+\infty}\mathcal{X}^\frac{3}{2}(r) \sqrt{r}\,dr},\quad  T = 0\,K.
	\end{cases}
\end{equation}

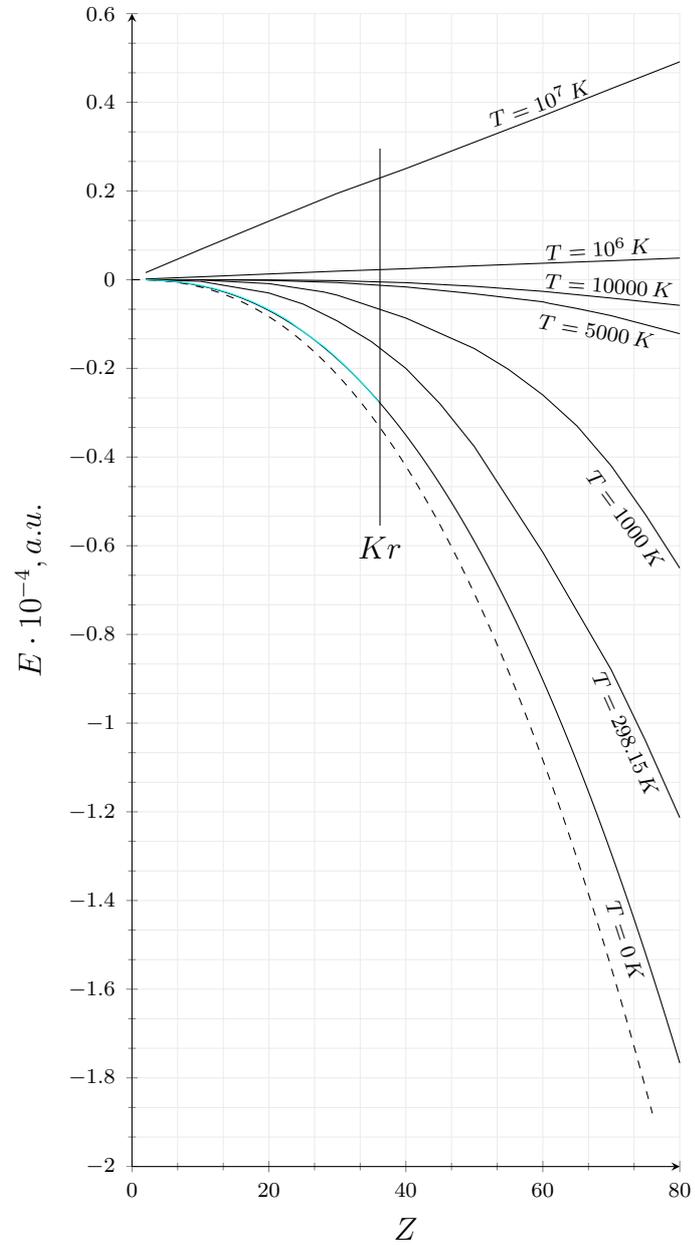
\begin{figure}
	\begin{center}
		\begin{tikzpicture}
			\begin{axis}[
	table/col sep = semicolon,
	width = 250pt, height = 480pt,
	xmin = 0,
	xmax = 80,
	ymax = 0.6,
	ymin = -2.0, 
	xlabel = {\normalsize $Z$}, 
	ylabel = {\normalsize $E \cdot 10^{-4},\,a.u.$}
	]
	\addplot [dashed, domain=0.0:76,samples=100, mark options={scale = 1.5}]
	{-0.00007687 *x^(7/3)};
	\addplot[mark options={scale = 1.5}] table [x={x}, y={y}] {ElementsT0K.csv} node[below, pos=0.93,yshift=0.12cm,rotate=-72]{\scriptsize $T=0\,K$};
	\addplot[mark options={scale = 1.5}] table [x={x}, y={y}] {ElementsT298.15K.csv} node[below, pos=0.93,yshift=0.1cm,xshift=0.0cm,rotate=-66]{\scriptsize$T=298.15\,K$};
	\addplot[mark options={scale = 1.5}] table [x={x}, y={y}] {ElementsT1000K.csv} node[below, pos=0.93,yshift=0.1cm,xshift=-0.08cm,rotate=-54]{\scriptsize$T=1000\,K$};		
	\addplot[mark options={scale = 1.5}] table [x={x}, y={y}] {ElementsT5000K.csv} node[below, pos=0.85,yshift=0.0cm,rotate=-10]{\scriptsize$T=5000\,K$};	
	\addplot[mark options={scale = 1.5}] table [x={x}, y={y}] {ElementsT10000K.csv} node[above, pos=0.85,yshift=-0.08cm,rotate=-4]{\scriptsize $T=10000\,K$};
	\addplot[mark options={scale = 1.5}] table [x={x}, y={y}] {ElementsT1000000K.csv} node[above, pos=0.85,yshift=-0.08cm,rotate=4]{\scriptsize $T=10^6\,K$};
	\addplot[mark options={scale = 1.5}] table [x={x}, y={y}] {ElementsT10.000.000K.csv} node[above, pos=0.75,yshift=-0.08cm,rotate=20]{\scriptsize $T=10^7\,K$};
	\addplot[cyan,mark options={scale = 1.5}] table [x={x}, y={y}] {CCSD.csv};									
\end{axis}
\draw [black](3.26, 8.5) node[below] {$Kr$} to (3.26, 13.5);
		\end{tikzpicture}
		\caption{\label{fig:4} The energy of electron gas in a field of atomic cores: black by $E=\mu Z$ for $Z=2 \div 80$, cyan by CCSD(T)/cc-pVQZ, dashed by $E=-0.7687 Z^\frac{7}{3}$ (Thomas-Fermi theory).}	
	\end{center}
\end{figure}
\begin{figure}
	\begin{center}
		\begin{tikzpicture}
			\begin{axis}[
				table/col sep = semicolon,
				width = 450pt, height = 250pt,
				xmin = 0,
				xmax = 490.0,
				ymin = -0.1,
				ymax = 0.1, 
				xlabel = {\normalsize $r, \; a.u.$}, 
				ylabel = {\normalsize $\varphi(r), \; a.u.$}
				]
				\addplot[mark options={scale = 1.5}] table [x={x}, y={y}] {He_5000.0K_phi.csv};
				\addplot [dashed, domain=0:200,samples=100, mark options={scale = 1.5}]
				{0.01063594219} node[right, pos=1.0,yshift=0.0cm]{$\mu - kT$};
				\addplot [dashed, domain=253:480,samples=100, mark options={scale = 1.5}]
				{0.01063594219};
			\end{axis}
		\end{tikzpicture}
		\caption{\label{fig:5} The electrostatic field near a helium core, $Z=2$, $T=5000\,K,\,\mu=0.02647\,a.u.$}	
	\end{center}
\end{figure}
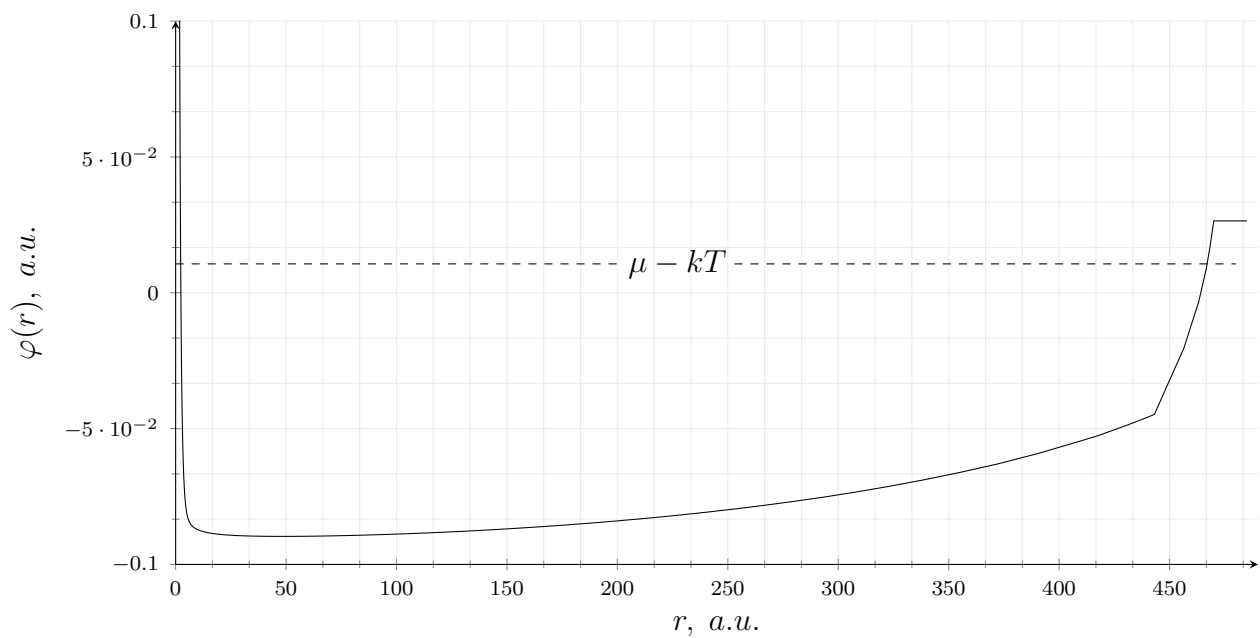
\begin{figure}
	\begin{center}
		\begin{tikzpicture}
			\begin{axis}[
				table/col sep = semicolon,
				width = 450pt, height = 250pt,
				xmin = 0,
				xmax = 490.0,
				ymin = -85.25,
				ymax = -85.15, 
				xlabel = {\normalsize $r, \; a.u.$}, 
				ylabel = {\normalsize $\varphi(r), \; a.u.$}
				]
				\addplot[mark options={scale = 1.5}] table [x={x}, y={y}] {atom_80_T_1000_K_phi.csv};
                 \addplot [dashed, domain=0:200,samples=100, mark options={scale = 1.5}]
{-85.20836681} node[right, pos=1.0,yshift=0.0cm]{$\mu - kT$};
	             \addplot [dashed, domain=253:470,samples=100, mark options={scale = 1.5}]
	{-85.20836681};
			\end{axis}
		\end{tikzpicture}
		\caption{\label{fig:6} The electrostatic field near a mercury core, $Z=80$, $T=1000\,K,\,\mu=-85.2052\,a.u.$}	
	\end{center}
\end{figure}
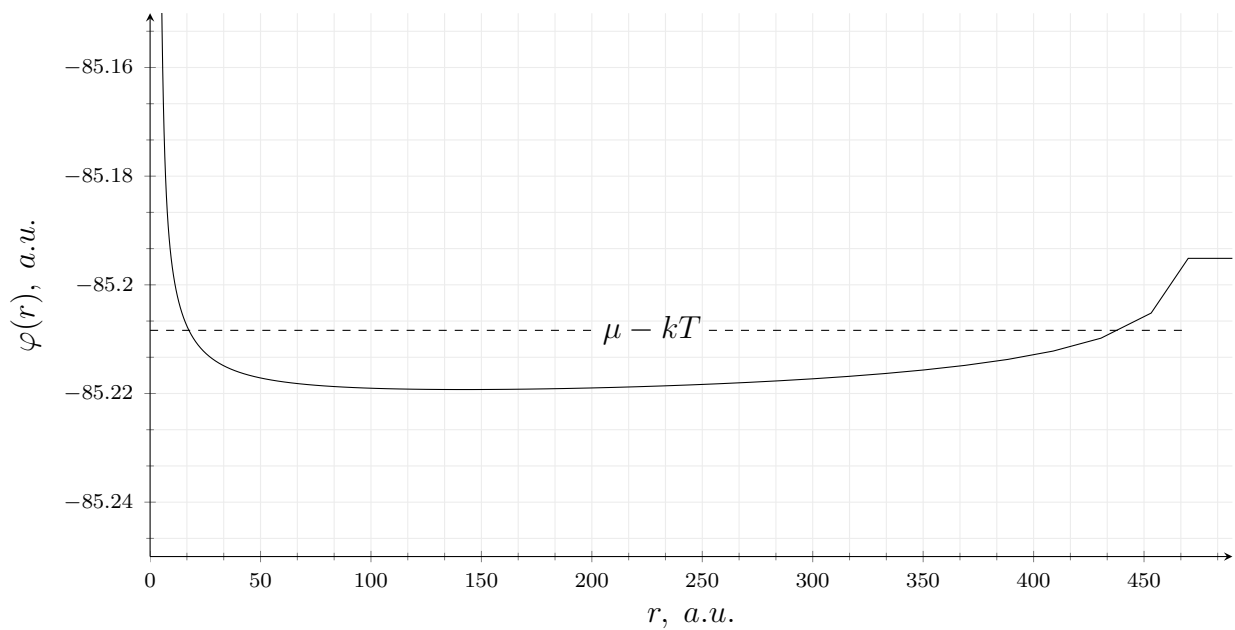

For $Z=1$ and $g=1$ the Eq. \eqref{eq:38} for $T=0\,K$ yields \cite{Kochnev2018} the energy of electron $E = 1 \cdot \mu_H$ in effective electrostatic field in the hydrogen atom $\mu_H \approx -\frac{1}{2}\,a.u.$ For $Z>1$ and $g=2$ for $T=0\,K$ it yields atomic energies $E=\mu Z$ which are in agreement with the Hartree-Fock calculations of atomic orbitals for light elements \cite{Kochnev2018,Kochnev_Chernogolovka,Kochnev2021}, Fig. \eqref{fig:4}, Tab. \eqref{E_table_T_0_K} in the appendix. The values $E=\mu Z$ at $T=0\,K$ for all atoms from He to $Z=128$ perfectly fit the power law of $7/3$ for atomic energies $E_{T=0\,K}(Z) = -0.640621 \cdot Z^\frac{7}{3}$.

Section of the state diagram of an inhomogeneous electron gas near atomic core for the range $Z = 2 \dots 80$ in Fig. \eqref{fig:4} shows the states of atoms with bounded electrons at low temperatures when the energy of electron density is negative, the state of perfect 'cold' plasma, when the energy of electron density is equal to zero, and the area of hot plasma when the energy of electron density is positive. 

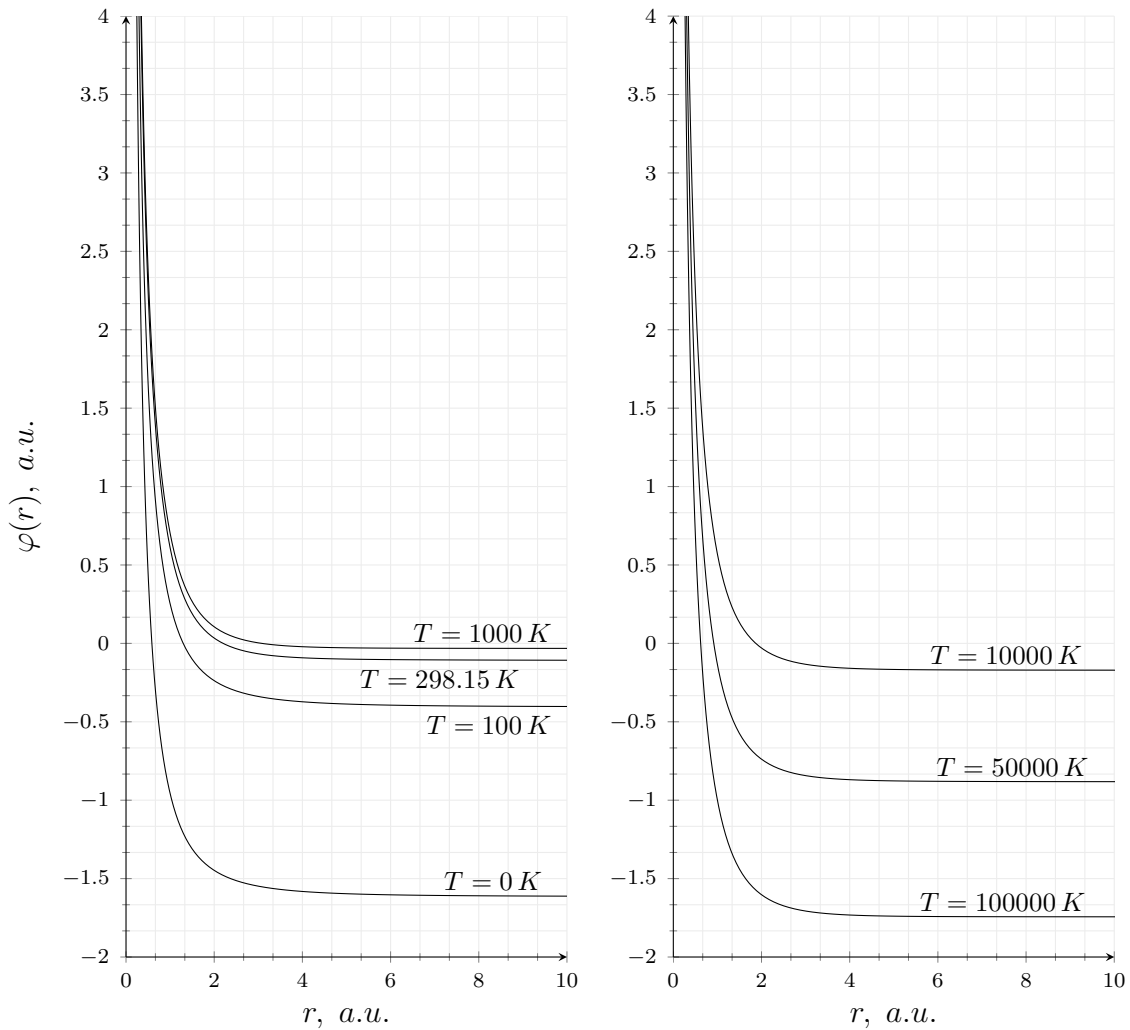
\begin{figure}
	\begin{center}
		\begin{tikzpicture}
			\begin{axis}[
				table/col sep = semicolon,
				width = 210pt, height = 400pt,
				xmin = 0,
				xmax = 10,
				ymin = -2.0,
				ymax = 4.0, 
				xlabel = {\normalsize $r, \; a.u.$}, 
				ylabel = {\normalsize $\varphi(r), \; a.u.$}
				]
				\addplot[mark options={scale = 1.5}] table [x={x}, y={y}] {He_0K_phi.csv} node[pos=0.70, yshift = 0.2cm,rotate=0] {\footnotesize  $T=0\,K$};
				\addplot[mark options={scale = 1.5}] table [x={x}, y={y}] {He_100.0K_phi.csv} node[pos=0.435, yshift = -0.25cm,rotate=0] {\footnotesize $T=100\,K$};				
				\addplot[mark options={scale = 1.5}] table [x={x}, y={y}] {He_298.15K_phi.csv} node[pos=0.67, yshift = -0.25cm,rotate=0] {\footnotesize $T=298.15\,K$};				
				\addplot[mark options={scale = 1.5}] table [x={x}, y={y}] {He_1000.0K_phi.csv} node[pos=0.67, yshift = 0.2cm,rotate=0] {\footnotesize $T=1000\,K$};	
			\end{axis}
		\end{tikzpicture}
		\begin{tikzpicture}
	\begin{axis}[
		table/col sep = semicolon,
		width = 210pt, height = 400pt,
		xmin = 0,
		xmax = 10,
		ymin = -2.0,
		ymax = 4.0, 
		xlabel = {\normalsize $r, \; a.u.$}
		]
		\addplot[mark options={scale = 1.5,restrict x to domain=0.0:10.0}] table [x={x}, y={y}]{He_10000.0K_phi.csv} node[pos=0.69, yshift = 0.2cm,rotate=0] {\footnotesize $T=10000\,K$};	
		\addplot[mark options={scale = 1.5,restrict x to domain=0.0:10.0}] table [x={x}, y={y}]{He_50000.0K_phi.csv} node[pos=0.69, yshift = 0.2cm,rotate=0] {\footnotesize $T=50000\,K$};			
       \addplot[mark options={scale = 1.5}] table [x={x}, y={y}] {He_100000.0K_phi.csv} node[pos=0.69, yshift = 0.2cm,rotate=0] {\footnotesize $T=100000\,K$};					
	\end{axis}
\end{tikzpicture}		
		\caption{\label{fig:7} The electrostatic field in vicinity of a helium atomic core.}	
	\end{center}
\end{figure}

For any positive temperature, heat phenomena in a gas of electrons are manifested through a temperature-dependent potential barrier between a small area near the atomic core and distant spatial points, as depicted in Figs. \eqref{fig:5}-\eqref{fig:6}, where the $\mu - kT$ levels are indicated\footnote{Since electrons are negatively charged, the barrier for them is convex downwards in Figs. \eqref{fig:5}-\eqref{fig:6}. The maximum radius $r_{max}=470\, a.u.$ was set when solving Eq. \eqref{eq:37}. Small computational artifacts at the right ends of computed curves are superficial and disappear when increasing the density of grid points or increasing $r_{max}$. These artifacts do not affect the field on the left side.}.

For positive temperatures, the behavior of atomic fields near the core is depicted in Fig. \eqref{fig:7} for helium. At $T=0\,K$, the field corresponds to a shielded Coulomb field of the core, with a negative absolute value of the chemical potential\footnote{Such a field is known as the <<effective>> potential in both Thomas-Fermi and Hartree-Fock theories;}. Negatively charged electrons are concentrated within the only region where $\varphi(r)$ has positive values, which is around the core. The chemical potential $\mu$ and the corresponding energy of the field increase monotonically with temperature, as shown in Table \eqref{tab:table2}. For sufficiently high temperatures, such as $T > 5000\,K$ for helium, the chemical potential becomes positive, and the field for any atomic core exhibits two regions with positive values of $\varphi(r)$. One is around the core, and the other has a minimum value of $\varphi = \mu$ at a distance from the core, beyond a temperature-dependent potential barrier, as shown in Figs. \eqref{fig:5}-\eqref{fig:6}. Near the core, the field retains the same shape for all temperatures but shifts vertically. The magnitude of this shift depends on both the chemical potential and the barrier.

As a result, for a helium atomic core, the field near the core initially moves upward in the temperature range of $T = 0 \dots 1000\,K$, remains nearly constant in the temperature range of $T = 1000 \dots 5000\,K$, and starts moving downward for $T > 5000\,K$, with an increasing height of the potential barrier that separates the vicinity of the atomic core from distant points. The motion of electrons in the field of any atomic core, at any temperature $T>0\,K$, can be described as the motion of particles between a potential well near the core and a potential valley at distanced points, separated by a potential barrier. The height of this barrier, in comparison to $kT$, is listed in Table \eqref{tab:table2}. For high temperatures and all atomic numbers, the positivity of $\mu$ implies that the location of electrons far away from the core is equally stable as the location near the core. However, at any temperature, the region near the atomic core is separated by a potential barrier from distant points, while the energy $kT$ is always lower than the height of the barrier.

 \renewcommand{\tablename}{Tab.}
\begin{table}[!h]
	\caption{\label{tab:table2} Temperature dependence of the electrostatic field in vicinity of a helium atomic core.}
	\begin{center}
		\begin{tabular}{|c||c|c|c|c|c|}
			\hline
			$T$,\,$K$ & $\mu$,\,$a.u.$ & Barrier from $\mu$, \,$a.u.$& $kT$,\,$a.u.$ & Min. of $\varphi(r)$,\,$a.u.$ & $E=\mu Z$,\,$a.u.$ \\
			\hline
			0 & $-1.61427$ & 0 & 0 & $-1.61427$ & $-3.22853$ \\
			100 & $-0.40442$ & $2.16567\cdot10^{-3}$ & $3.16681\cdot10^{-4}$ & $-0.40659$ & $-0.80884$ \\	
			298.15 & $-0.10483$  & $6.84523\cdot10^{-3}$ & $9.44185\cdot10^{-4}$ & $-0.11168$ & $-0.20966$ \\
			1000 & $-0.01421$  & $1.83347\cdot10^{-2}$ & $3.16681\cdot10^{-3}$ & $-0.03255$ & $-0.02843$ \\
			5000 & 0.02647 & 0.11613 & $1.58341\cdot10^{-2}$ & $-0.08965$ & 0.05294 \\
			10000 &  0.06030 & 0.11171 & $3.16681\cdot10^{-2}$ & $-0.17201$ & 0.12059 \\
			50000 & 0.33094 & 0.55165 & $1.58341\cdot10^{-1}$ & $-0.88258$ & 0.66187 \\
			$10^5$ &  0.68484 & 1.05891 & $3.16681\cdot10^{-1}$ & $-1.74375$ & 1.36968 \\
			$10^6$ &  7.53096 & 11.2616 & $3.16681$ & $-18.7925$ & 15.0619 \\
			\hline
		\end{tabular}
	\end{center}
\end{table}

\section{Many-body subsystem}

Bodies with negligible mass compared to the mass of the hot gravitational source or the apparent mass of the gas surrounding the source can be described without the need to introduce the mechanical equivalent of heat. Instead, they move mechanically within the potential field, taking into account all heat phenomena present in the field. Achieving homogeneity of degree one with respect to the number of particles then results in mechanical shifts of partial values of the class representative functionals, where the shifts reflect interactions between particles.

The motion of material points on the time segment \([t_0, t_1]\) is described by those functions \(x_i(t)\), \(y_i(t)\), \(z_i(t)\) that give a stationary value to the integral of action \(S\), introduced in the footnote \eqref{footnote_29}. When the functional \(S[x_i(t)\), \(y_i(t)\), \(z_i(t)]\) reaches a stationary value, the equations \(\frac{\delta S}{\delta x_i} = \frac{\partial L}{\partial x_i} - \frac{d}{dt}\frac{\partial L}{\partial \dot{x_i}} = 0\) must be satisfied. Trajectories \(x_i(t)\), \(y_i(t)\), \(z_i(t)\) can be found from these equations for a given initial values \(x_i(t_0)\), \(y_i(t_0)\), \(z_i(t_0)\) and \(\dot{x_i}(t_0)\), \(\dot{y_i}(t_0)\), \(\dot{z_i}(t_0)\). The action \(S\) is homogeneous of order one regarding masses \(m_i\), and the Eqs. \eqref{eq:homogeneity} and \eqref{eq:chem_potentials} are not restricted to the deterministic trajectories; the stationarity of \(S\) is applicable to statistically stationary random processes in the movement of the particles. Trajectories of particles now are \(x_i(t) + X_i(t)\), \(y_i(t) + Y_i(t)\), \(z_i(t)+Z_i(t)\), where \(x_i(t)\), \(y_i(t)\), \(z_i(t)\) are solutions of the motion equations \(\frac{\delta S}{\delta x_i} = 0\), and \(X_i(t)\), \(Y_i(t)\), \(Z_i(t)\) are trajectories of statistically stationary random processes with zero mean values\footnote{here is supposed the variation functional $\delta S[x_i(t),\,y_i(t),\,z_i(t), \delta x_i(t),\, \delta y_i(t),\, \delta z_i(t)]$ is a continuous by $x_i(t),\,y_i(t),\,z_i(t)$, thus, for $\delta S[x_i(t) + X_i(t),\,y_i(t) + Y_i(t),\,z_i(t)+Z_i(t), \delta x_i(t),\, \delta y_i(t),\, \delta z_i(t)]$ its expected value $E\delta S[x_i(t) + X_i(t),\,y_i(t) + Y_i(t),\,z_i(t)+Z_i(t), \delta x_i(t),\, \delta y_i(t),\, \delta z_i(t)]=\delta S[x_i(t) + EX_i(t),\,y_i(t) + EY_i(t),\,z_i(t)+EZ_i(t), \delta x_i(t),\, \delta y_i(t),\, \delta z_i(t)]=\delta S[x_i(t),\,y_i(t),\,z_i(t), \delta x_i(t),\, \delta y_i(t),\, \delta z_i(t)]$ is the variation of $S$;}. For the statistical stationarity, the processes have spectral representations on a time interval \(T=t_1-t_0\) by Fourier series \(X_i(t)=\sum_{n=-\infty}^{\infty}\,c_n e^{iw_nt}\) with random coefficients \(c_n=a_n+ib_n\), wherein \(c_0 = 0\), \(E[c_n] = 0\), and \(E[c_ic_j] = 0\) if \(i \neq j\). Frequencies \(w_n=\frac{2\pi n}{T}\) are defined for all integer \(n\), and \(w_{-n}=-w_n\). For real-valued fluctuations, conjugate symmetry \(c_n=c_{-n}^*\) gives \(X_i(t)=2\sum_{n=1}^{\infty}\,[a_n \cos w_nt-b_n\sin w_nt]\). The spectral representation of the process defines representations of its time derivatives by differentiation of the series. It is required to specify the probability distributions of random numbers \(a_n\) and \(b_n\). The velocities \(\dot X_i(t)\), \(\dot Y_i(t)\),  \(\dot Z_i(t)\) can be discontinuous due to collisions of particles, but increments of the trajectories \(X(t)-X(t_0)= \int_{t_0}^{t}\, \dot X_i(t) dt\) are absolutely continuous.

The time average of energy \(\langle H \rangle_T = \frac{1}{T} \int_{t_0}^{t_1} H dt\) over the period is \(\langle H \rangle_T = \sum_i \bigg\langle \dfrac{\partial H}{\partial m_i}\bigg\rangle_T m_i\). Suppose, there are \(N\) bodies observed, and they interact by the universal law of gravitation with the interaction energy \(U=- G\sum_{i < j} \frac{m_im_j}{r_{ij}}\). The partial values of the time average of energy have kinetic and potential contributions\footnote{$\mu_i=\big\langle \frac{\partial H}{\partial m_i} \big\rangle_T =\big\langle \frac{1}{2} \big((\dot{x_i} + \dot{X_i})^2+(\dot{y_i} + \dot{Y_i})^2 + (\dot{z_i} + \dot{Z_i})^2\big) \big\rangle_T - \big\langle G \frac{\partial}{\partial m_i} \sum_{i < j} \frac{m_im_j}{r_{ij}} \big\rangle_T$, where $\big\langle \frac{1}{2} (\dot{x_i} + \dot{X_i})^2 \big\rangle_T = \big\langle \frac{1}{2}\dot{x_i}^2 \big\rangle_T + 0 + \frac{1}{2}D\dot{X_i}$ and similarly for $\big\langle \frac{1}{2} (\dot{y_i} + \dot{Y_i})^2 \big\rangle_T$, $\big\langle \frac{1}{2} (\dot{z_i} + \dot{Z_i})^2 \big\rangle_T$ and $\frac{\partial}{\partial m_i} \sum_{i < j} \frac{m_im_j}{r_{ij}} = \sum_{j \ne i} \frac{m_j}{r_{ij}}$;}:
\begin{equation}\label{eq:39}
	\mu_i = \bigg\langle \dfrac{\partial H}{\partial m_i}\bigg\rangle_T = \bigg\langle \frac{1}{2}\big( \dot x_i^2 + \dot y_i^2 + \dot z_i^2 \big)\bigg\rangle_T + \frac{3}{2}  D \dot X_i - u_i.
\end{equation}
Kinetic contribution of variances of velocities $\dot X_i(t)$, $\dot Y_i(t)$,  $ \dot Z_i(t)$ is denoted $\frac{3}{2} D \dot X_i$, and interaction contribution is denoted $ \bigg\langle G\,\sum_{j \ne i} \frac{m_j}{r_{ij}} \bigg\rangle_T = u_i$. Wherein all complicated interactions result in a shift $u_i$ of the value of $i$-th chemical potential. The energy is $\langle H \rangle_T = \sum_i\, \mu_i\, m_i$ and action per period is $S / T = \sum_i\, ( \big\langle \frac{1}{2}\big( \dot x_i^2 + \dot y_i^2 + \dot z_i^2 \big)\big\rangle_T + \frac{3}{2} D \dot X_i + u_i )\, m_i$ where the partial value is shifted for $u_i$ due to interactions between bodies.

For a given initial values \(x_i(t_0)\), \(y_i(t_0)\), \(z_i(t_0)\) and \(\dot{x_i}(t_0)\), \(\dot{y_i}(t_0)\), \(\dot{z_i}(t_0)\), and for a given probability distributions of independent random coefficients \(a_i\) and \(b_i\), the motion of particles is completely described in terms of the stationary action. If it is required to have predetermined positions at \(t_0\) and \(t_1\), the dropping cosines in the Fourier series and defining suitable random coefficients \(b_i\) for sines implement the motion of a particle along a random or randomized trajectory\footnote{the expected value of energy for the randomization of trajectories can be found by the virial theorem \(2\langle T \rangle_T = -\langle U \rangle_T\) for interacting bodies, which holds also for Fourier series with random coefficients; this allows finally writing out the energy $\langle H \rangle_T = \langle T \rangle_T + \langle U \rangle_T = -\langle T \rangle_T$ and minimal action for the period $S / T =\langle T \rangle_T - \langle U \rangle_T = 3\langle T \rangle_T$; wherein $\langle T \rangle_T = \sum_i \frac{m_i}{2}\int_{t_0}^{t_1}(\dot x_i^2 + \dot y_i^2 + \dot z_i^2)dt + \frac{3}{2} \sum_i m_i D \dot X_i$ consists of the kinetic energy for deterministic smooth motion \(x_i(t)\), \(y_i(t)\), \(z_i(t)\) of material points, and kinetic energy of stochastic deviations \(X_i(t)\), \(Y_i(t)\), \(Z_i(t)\); as long as the energy and the action are expressed through kinetic energy in the manner like the interacting bodies do not interact, it is seen, that the average energy consumption for a random contribution of the motion of $i$-th body is  \(m_i \cdot \frac{3}{2} D \dot X_i\); } that will start from a given point at \(t_0\) and arrive at a predetermined destination point at exactly a predetermined time \(t_1\).
 
Observing the interaction between particles, one immediately gets a functional description of their motion as if they were noninteracting particles, where all possible interactions are accounted for in the shifts \(u_i\) in Eq. \eqref{eq:39}.

\section{Discussion}

The phenomena under consideration are related to the richness of classes of equivalent functionals used to describe the equilibrium of the system of particles and the motion of particles. As demonstrated by variational arguments \eqref{eq:homogeneity} and \eqref{eq:chem_potentials}, each class of equivalent functionals, which have the same or equal Euler's equations for states of the system, contains a representative functional that is homogeneous of degree one in the number of particles in the system. This fact removes any differences in consideration, whether it's regarding interacting or non-interacting particles and bodies, large or small systems, known explicit energy/action functionals, or unknown ones. It always relies on the predefined structure and properties of homogeneous functions of degree one.

Expressions of classical Thermodynamics rely on Euler's integration of homogeneous functions.
If $Y(x_1,\,x_2,\,\vec{n})$ is an extensive property depending on intensive variables $x_1,\,x_2$ and a vector of extensive variables $\vec{n}$, the total differential

\begin{equation}
	dY = \bigg(\frac{\partial Y}{\partial x_1}\bigg)_{x_2,\,\vec{n}}dx_1 + \bigg(\frac{\partial Y}{\partial x_2}\bigg)_{x_1,\,\vec{n}}dx_2 + \sum_{i=1}^{K} \bigg(\frac{\partial Y}{\partial n_i}\bigg)_{x_1,\,x_2,\,\vec{n'}}dn_i
\end{equation}
gets integrated to $Y=\sum_{i=1}^{K} \mu_i n_i$ for $\mu_i=\left(\frac{\partial Y}{\partial n_i}\right)_{x_1,\,x_2,\,\vec{n'}}$ (when integrating, intensive function arguments are ignored). Thus, the internal energy total differential $dU=TdS-pdV+\mu dN$ gets integrated to $U=TS-pV+\mu N$, while the Gibbs energy total differential $dG=-SdT+Vdp+\mu dN$ gets integrated to $G=\mu N$.  As long as by definition $G=U+pV-TS$, the integrals are equal up to the shift. Relying on the variational arguments \eqref{eq:homogeneity} and \eqref{eq:chem_potentials} the Euler's integration can be conducted for a general, not necessarily an extensive property of a system of particles, writing down the homogeneous functional defined up to the equivalence, as detailed in section 4 above. The total differential of energy $E=E(N, \upsilon)$ of $N$ electrons \cite{Parr1978}\footnote{See Eq. (10) in the section 'DENSITY FUNCTIONAL THEORY' in \cite{Parr1978};} in the outer potential $\upsilon$

\begin{equation}
	dE = \mu dN - \int \rho\, d \upsilon(1)\, dV
\end{equation}
 gets integrated to $E=\mu N$.

The smallest physical system is an elementary cell $h^3$ of phase space of coordinates and momenta. This smallest system possesses its temperature $T$, entropy $S$, and a specific value of the chemical potential $\mu$ if there are occupying particles. Statistical properties of this smallest system are described in terms of the canonical ensemble, leading to the emergence of Fermi-Dirac and Bose-Einstein statistics. The presence of an outer reservoir with a large number of particles is redundant, and the exact number of particles or its definiteness is not significant.

The gradient of the chemical potential for particles can generate forces acting on particles in addition to forces produced by force fields, as illustrated in Fig. \eqref{fig:dV}. Applying mechanical laws to this situation allows us to derive mass transfer equations for ideal systems. However, when the laws for ideal systems are not applicable, or a well-defined functional for investigating equilibrium is unknown, algebraic arguments based on homogeneous functions of degree one can be employed. In this approach, the system of interest is represented as an assembly of elementary cells $h^3$ in phase space. The phase-space-based density is defined using statistical methods, and the invariant Gauss theorem is utilized to determine the resulting potential field as a solution of the Poisson differential equation. This approach enables the description of the physical system in local detail, independently of its overall view. As long as the exclusion principle is applied to the particles within the cell $h^3$, the phase-space-based density cannot be normalized to a specific number of particles. However, it can be scaled down without violating the exclusion principle, using a factor less than one, resulting in a rarefied gas in the phase space of coordinates and momenta. To normalize the phase-space-based density of particles to a given number of particles, an appropriate value for the chemical potential $\mu$ must be assigned.

The distinction between the phase-space-based density 'in small' and a spatial number density, defined as the number of particles per unit volume, leads to a peculiar illusion at high positive temperatures. This illusion can create the appearance of additional mass or charge because the apparent mass or charge related to the phase-space-based density can exceed the real mass corresponding to the value of the chemical potential. The temperature-dependent field illustrated in Figs. \eqref{fig:2}-\eqref{fig:3} causes a significant deviation in motion from Newton's law of universal gravitation without the need to postulate any invisible mass. This effect can be explained simply by considering that the visible mass is hot and has a non-zero temperature $T > 0\,K$. The deviation from the law of universal gravitation is then attributed to heat phenomena in the gravitational field, while Newton's law of universal gravitation itself arises from the partition function applied to the dense source of gravitation, based on fundamental physical constants.

The spatial density of particles is the integral
\begin{equation}\label{eq:quantum_mechanical_density}
	\rho(1)=N \int \left\| \Psi(1,\,w_1,\,2,\,w_2 \dots,\,N,\,w_N ) \right\|^2\,dw_1\,dx_2\,dx_3 \dots dx_N,
\end{equation}
where $dx_i = dw_i dV_i$ is a space-spin volume element, with $dw_i$ the spin part, and $\Psi(1,\,w_1,\,2,\,w_2 \dots,\,N,\,w_N )$ is a wave-function of $N$ particles depending on $N$ space-spin tuples $(x_i,\,y_i,\,z_i,\,w_i)$ of particles' coordinates. The density \eqref{eq:quantum_mechanical_density} is spin free\footnote{for this density \textit{real} number of particles is $N=N\left[ \rho \right]=\int \rho dV$;}.

The distinction between the phase-space-based number density \eqref{eq:25} and a spatial number density \eqref{eq:quantum_mechanical_density} arises from the application of the continuity equation $\frac{\partial \rho}{\partial t} + \nabla \cdot \vec{j} = 0$ to distinct flux vectors. Equations \eqref{eq:thermal_effects} pertain to the flux of particles \eqref{eq:flux}, while Equations \eqref{eq:30}, \eqref{eq:33}, and \eqref{eq:37} pertain to fluxes that differ from \eqref{eq:flux} due to their heat components, influenced by the relevant equations of state for particles under different physical conditions. The density \eqref{eq:quantum_mechanical_density} is related to the flux 
\begin{equation}\label{eq:quantum_flux}
		\vec{j}(\vec{r}, t)= \frac{h}{4 \pi m i}\left(\psi^* \nabla \psi - \psi \nabla \psi^* \right)
\end{equation}
of probability density, where $i^2=-1$, and $\psi(\vec{r}, t)$ represents the wave function of a single particle\footnote{This expression for the flux of probability density is well-known in quantum mechanics for a single particle in a potential field, non-interacting with other particles. Here, as any functional for particles has a homogeneous equivalent of degree one, any system of particles can be accurately described as a system of non-interacting particles moving in an effective potential that accounts for all interactions. Consequently, the flux expression applies to particles in a system of interacting particles as well.} with mass $m$ within a system of particles described by the wave function $\Psi$. This leads to different behavior near asymptotic boundary conditions. The density corresponding to the flux of type \eqref{eq:flux} becomes unbounded near the central region $\varphi\sim \frac{Z}{r}$ of the force field for an unbounded force. In contrast, the density \eqref{eq:quantum_mechanical_density} for the flux \eqref{eq:quantum_flux} remains finite at $r=0$, which is related to the uncertainty relations.

The energy expression for densities corresponding to the flux of type \eqref{eq:flux} is a homogeneous integral functional of degree one: $\int m \rho \varphi dV$ for mass density in a gravitational field and $\int (-1) \rho \varphi dV$ for electrons in an electric field in atomic units. These expressions can be represented algebraically as $E = \mu N$. In the case of gravitation, the number of particles $N = \frac{M_{gas}}{m}$ in the expression \eqref{eq:35} for the chemical potential can be assigned any value for mass $M_{gas}$. However, for electrons in the field of an atomic core with charge symmetry throughout the entire space, the number of electrons must be equal to the atomic number $Z$ in the expression \eqref{eq:38} for the chemical potential. The depiction of heat phenomena in atomic electric fields involves the development of temperature-dependent barriers between the center and distant points, as shown in Figs. \eqref{fig:5} and \eqref{fig:6}. Even though a non-neutral space might be produced with $N \neq Z$ in \eqref{eq:38}, the appearance of temperature-dependent barriers remains consistent due to the properties of the $F_\frac{1}{2}$ function.

The energy of the spatial density \eqref{eq:quantum_mechanical_density} is given by
\begin{equation}
	\left\langle \Psi | \hat{H} \Psi \right\rangle = \int \Psi^*(1,\,w_1,\,2,\,w_2 \dots,\,N,\,w_N ) \hat{H}\Psi(1,\,w_1,\,2,\,w_2 \dots,\,N,\,w_N )dx_1\,dx_2\,dx_3 \dots dx_N,
\end{equation}
where the operator of energy $\hat{H}$ acting on the wave function $\Psi$ represents the procedure of measuring the energy of the system of particles. The functional $\left\langle \Psi | \hat{H} \Psi \right\rangle$ belongs to a class of equivalent functionals $\mathcal{F} / \left\langle \Psi | \hat{H} \Psi \right\rangle$ that have the same or equal Euler's equations for $\Psi$, where $\mathcal{F}$ stands for functionals acting on $\Psi \in L_2$. According to the homogeneity argument \eqref{eq:homogeneity}, this class $\mathcal{F} / \left\langle \Psi | \hat{H} \Psi \right\rangle$ contains a representative functional $\left\langle \Psi | \hat{H}_0 \Psi \right\rangle$ that is homogeneous of degree one with respect to the number of particles. Quantum mechanics requires the operator $\hat{H}_0$ to be hermitian. Thus, $\hat{H}_0 = \sum_i \hat{h}(i)$ is a sum of one-electron operators, and the wave function $\Psi$ can be represented through one-electron orbitals describing the states of electrons in potentials depicted in Figs. \eqref{fig:5}, \eqref{fig:6}, \eqref{fig:7}. Any changes in the state of the atom at a positive temperature are directly represented by the movement of particles through a potential barrier, while at zero temperature, the representation is similar to the Hartree-Fock theory with an effective potential.

A particular problem for functionals acting on particle number density is the Hohenberg-Kohn theorem \cite{Hohenberg1964}. This theorem establishes a one-to-one correspondence\footnote{The Hohenberg-Kohn theorem states that for any system of interacting particles in an external potential $v(\vec{r})$, the external potential is uniquely determined (except for a constant) by the ground state density $n_0(\vec{r})$.} between the ground state spatial density of electrons, as described by \eqref{eq:quantum_mechanical_density} at $T=0\,K$, and the external potential $v$. Numerous suggestions have been made for the functional of particle density, with one of the most notable being the B3LYP functional \cite{Lee1988}. Parr postulated\footnote{See equation (30) in the section 'POSSIBLE HOMOGENEITIES' in \cite{Parr1978}.} that the energy of electron density is given by $E = \mu N - \sum_k (\varkappa_k - 1)f_k$, where $f_k[\rho]$ are functionals homogeneous of degree $\varkappa_k$ with respect to $\rho$. According to the homogeneity equation \eqref{eq:homogeneity}, it is always possible to find a representation in which the latter expression is exactly $E = \mu N$. All assumptions about the form of energy functionals for particle density ultimately converge, to varying degrees, toward the functional $E\left[N,\,\rho(1)\right] = \int \rho(1) \nabla^{-2}\rho(1) dV$, where, in Cartesian coordinates, $\nabla^2$ is the Laplace operator. The application of this functional involves two or three steps: (a) solving the Poisson equation with phase-space-based density on the right-hand side, (b) assigning an appropriate value to the chemical potential $\mu$ for the normalization of the phase-space-based density to a given number of particles and computing the energy as $E = \mu N$, and (c) if the spatial density and all other properties of the particle system are of interest, finding a wave function for non-interacting particles in the outer potential determined in (a) and (b).

\section{Conclusion}

Heat phenomena in potential fields occur at any positive temperature, regardless of the scale of the physical system of particles.

\pagebreak

\section{Appendix}

\begin{figure}
	\begin{center}
		\begin{tikzpicture}
			\pgfplotsset{
				log x ticks with fixed point/.style={
					xticklabel={
						\pgfkeys{/pgf/fpu=true}
						\pgfmathparse{exp(\tick)}%
						\pgfmathprintnumber[fixed relative, precision=3]{\pgfmathresult}
						\pgfkeys{/pgf/fpu=false}
					}
				},
				log y ticks with fixed point/.style={
					yticklabel={
						\pgfkeys{/pgf/fpu=true}
						\pgfmathparse{exp(\tick)}%
						\pgfmathprintnumber[fixed relative, precision=3]{\pgfmathresult}
						\pgfkeys{/pgf/fpu=false}
					}
				}
			}
			
			\begin{axis}
				[
				table/col sep = semicolon,
				xmode=log,
				ymode=log,
				log x ticks with fixed point,
				axis x line=bottom,
				axis y line=left,
				ymax = 1.20871*10^10,
				xlabel={ \normalsize $\eta$ },
				ylabel={\normalsize $ F_{\frac{1}{2}}(\eta)$ },
				every axis x label/.style={
					at={(ticklabel* cs:1.05)},
					anchor=west,
				},
				every axis y label/.style={
					at={(ticklabel* cs:1.05)},
					anchor=south,
				},
				ytick={10, 100, 1000, 10000, 100000, 1000000, 10000000, 100000000, 1000000000},  
				scale=1.6,
				grid=both,
				tick align=outside,
				tickpos=left,
				]
				\addplot[mark options={scale = 1.5}] table [x={x}, y={y}]{F05.csv};
			\end{axis}
		\end{tikzpicture}
		\caption{\label{fig:F05} Fermi-Dirac function of order $\frac{1}{2}$.}
	\end{center}
\end{figure}
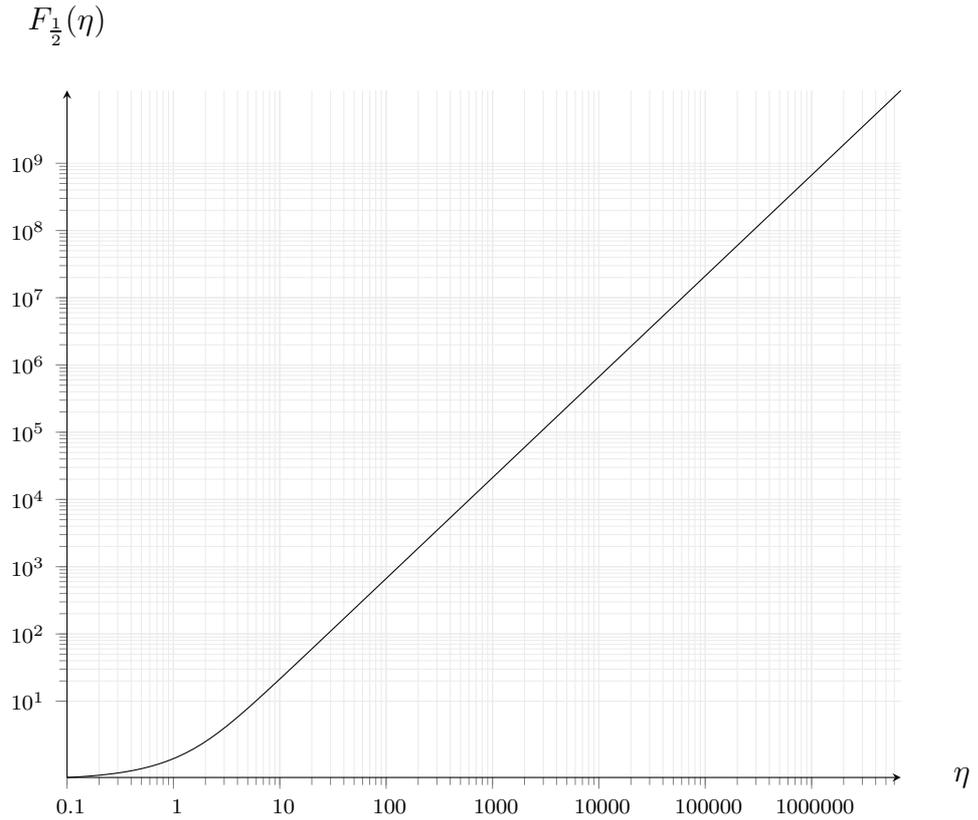

The graph of the special Fermi-Dirac function of order $\frac{1}{2}$ is depicted in Fig. \eqref{fig:F05}. The function $F_{\frac{1}{2}}$ exhibits an exponential-like behavior for large arguments, but its derivative $F'_{\frac{1}{2}}$ deviates from exponential growth and eventually decays as $\eta$ becomes sufficiently large.\\

Functional for the Eq.\eqref{eq:33} for $T>0\,K$ reads
\begin{equation}\label{eq:I1}
	I_1[\mathcal{X}(r)] = \int\limits_{0}^{+\infty}\Biggl\{ \frac{1}{2} \Bigg( \frac{d\mathcal{X}}{dr} \Bigg)^2 -  \frac{1}{\pi} \,c \, \sqrt{2} (kT)^\frac{5}{2}\cdot r^2 \int_{0}^{-4\pi^2m\mathcal{X}(r) / (r\,kT)} F_{\frac{1}{2}}(\eta)\, d\eta \Biggr\} \, dr,
\end{equation}
and when $T=0\,K$ reads
\begin{equation}\label{eq:I2}
	I_2[\mathcal{X}(r)] = \int\limits_{0}^{+\infty}\Biggl\{ \frac{1}{2} \Bigg( \frac{d\mathcal{X}}{dr} \Bigg)^2 - \frac{16 \pi}{15} \, c \, G^\frac{3}{2} m^\frac{5}{2} \; \frac{1}{\sqrt{r}} \; (\mathcal{-X})^\frac{5}{2} \Biggr\} \, dr.
\end{equation}

Functional for the Eq.\eqref{eq:37} for $T>0\,K$ reads
\begin{multline}\label{eq:I3}
	I_3[\mathcal{X}(r)] = \int\limits_{0}^{+\infty}\Biggl\{ \frac{1}{2} \Bigg( \frac{d\mathcal{X}}{dr} \Bigg)^2 + \frac{2g }{\pi} \cdot r \int_{0}^{\mathcal{X}} d\mathcal{X} \cdot \int_{0}^{+\infty} \frac{p^2 dp}{e^{\big(\frac{p^2}{2}-\frac{\mathcal{X}}{r}\big)\cdot \frac{1}{kT}}+1} \Biggr\} \, dr =\\
	\int\limits_{0}^{+\infty}\Biggl\{ \frac{1}{2} \Bigg( \frac{d\mathcal{X}}{dr} \Bigg)^2 + \frac{2g \sqrt{2}}{\pi} (kT)^\frac{5}{2}\cdot r^2 \int_{0}^{\mathcal{X}(r) / (r\,kT)} F_{\frac{1}{2}}(\eta)\, d\eta \Biggr\} \, dr,
\end{multline}
and when $T=0\,K$ reads
\begin{equation}\label{eq:I4}
	I_4[\mathcal{X}(r)] = \int\limits_{0}^{+\infty}\Biggl\{ \frac{1}{2} \Bigg( \frac{d\mathcal{X}}{dr} \Bigg)^2 + \frac{8g \sqrt{2}}{15\pi} \; \frac{1}{\sqrt{r}} \; \mathcal{X}^\frac{5}{2} \Biggr\} \, dr.
\end{equation}

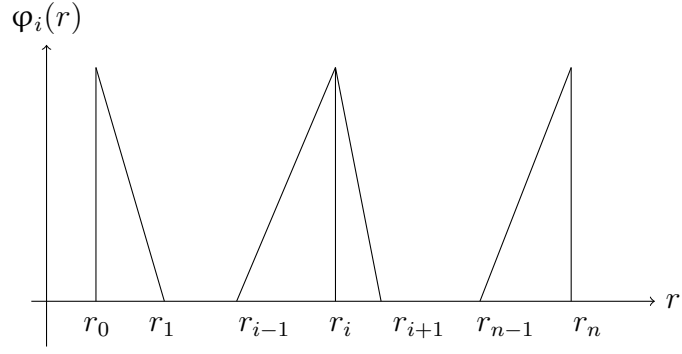
\begin{figure}
	\begin{center}
		\begin{tikzpicture}[domain=0:6] 
			\draw[->] (-0.2,0) -- node[below=2pt] {$  r_0 \quad \:r_1 \qquad r_{i-1} \:\quad r_i \;\quad r_{i+1} \quad r_{n-1} \quad \: r_n$} (8.0,0) node[right] {$r$}; 
			\draw[->] (0,-0.6) -- (0,3.4) node[above] {$\upvarphi_i(r)$};
			\draw (0.65, 0) to (0.65, 3.1);
			\draw (0.65, 3.1) to (1.55, 0);
			\draw (2.5, 0) to ( 3.8, 3.1 );
			\draw ( 3.8, 3.1 ) to (3.8, 0);
			\draw ( 3.8, 3.1 ) to (4.4, 0);
			\draw ( 5.7, 0 ) to (6.9, 3.1);
			\draw (6.9, 3.1) to (6.9, 0);
		\end{tikzpicture}
		\caption{\label{fig:basis} Basis functions.}
	\end{center}
\end{figure}

In the Euler broken line method an unknown function is approximated piece-wise linearly $\mathcal{X}(r) \approx \mathcal{X}_1(r)$ with respect to the values at nodes,  $\mathcal{X}_1(r) = \mathcal{X}_{k-1} \cdot \frac{r_k - r}{r_k -r_{k-1}}+\mathcal{X}_k \cdot \frac{r-r_{k-1}}{r_k-r_{k-1}}$ for $r \in [r_{k-1},\,r_k]$, that is expanded by pyramidal basis functions, Eq. \eqref{eq:basis}, Fig. \eqref{fig:basis}, as $\mathcal{X}_1(r) = \sum_{i=0}^{n} \varphi_i(r)\,\mathcal{X}_i$ where the coefficients before basis functions are the values of the unknown function $\mathcal{X}_i$ at grid points $r_i$.

\begin{multline}\label{eq:basis}
	\varphi_0(r)=
	\begin{cases}
		\frac{r_1-r}{r_1-r_0} \qquad (r_0 \le r \le r_1)
		\\
		0 \qquad \qquad   (r_1 \le r \le r_n),
	\end{cases}
	\\
	\varphi_i(r)=
	\begin{cases}
		0 \qquad \qquad (0 \le r \le r_{i-1})
		\\
		\frac{r - r_{i-1}}{r_i-r_{i-1}} \qquad (r_{i-1} \le r \le r_i)
		\\
		\frac{r_{i+1}-r}{r_{i+1}-r_i} \qquad (r_i \le r \le r_{i+1})
		\\
		0 \qquad \qquad (r_{i+1} \le r \le r_n), \qquad \qquad \qquad \qquad
	\end{cases}
	\\
	\varphi_n(r)=
	\begin{cases}
		0 \qquad  \qquad  (0 \le r \le r_{n-1})
		\\
		\frac{r-r_{n-1}}{r_n-r_{n-1}} \qquad (r_{n-1} \le r \le r_n).\qquad \qquad \qquad 
	\end{cases}
\end{multline}

This involves a continuous approximation, and it is well-known that the approximation converges uniformly to a solution obtained using the finite element method. We can construct a piece-wise linear approximation of the solution by applying the finite element method and employing the Ritz method based on the corresponding variational principle $\delta I[\mathcal{X}] = 0$. Let's choose a sufficiently large value $r_{\text{max}}$ and divide the interval $[0, r_{\text{max}}]$ into points $r_k$. The stationarity condition for the functional $I[\mathcal{X}(r)]$ leads to a system of algebraic equations:
\begin{equation}\label{eq:48}
	\frac{\partial}{\partial \mathcal{X}_k }\,I\Bigg[ \sum_{i=0}^{n} \varphi_i(r)\,\mathcal{X}_i \Bigg] = 0 \qquad (k = 1, \dots, n - 1),
\end{equation}
where boundary values $\mathcal{X}_0 = -M_{\odot}$ or $\mathcal{X}_0 = Z$ are prescribed, and $\mathcal{X}_N = 0$ is enforced.\\

The system of equations \eqref{eq:48} is solved using a safe version of the Newton-Raphson method with a check for a global convergence failure event. After each solution, the density of points $r_i$ is doubled, and the iteration is repeated until the desired tolerance between solutions is achieved. To facilitate this process, a large precomputed table of $F_{\frac{1}{2}}(\eta)$ function values and their derivatives is required for Hermitian interpolation. Subsequently, the derivatives \eqref{eq:48} and all underlying derivatives in the Newton-Raphson procedure are computed using automatic differentiation to ensure machine accuracy. While simple finite differences can be applied to Eqs. \eqref{eq:33} and \eqref{eq:37}, this approach suffers from larger variations and slower convergence between iterations. If an effective means for computing the $F_{\frac{1}{2}}$ function is not available, the schemes \eqref{eq:I1} and \eqref{eq:I3} naturally cancel out errors in $F_{\frac{1}{2}}$ due to its crude computation, as $F_{\frac{1}{2}}(\eta)$ is integrated under the integral.

\vskip 2cm

\begin{longtable}[c]{| c | c | c | c | c |c|}
	\caption{Energies of Elements at $T=0\,K$.\label{E_table_T_0_K}}\\
	\hline
	$Z$ & Element & $\mu,\,a.u.$, \eqref{eq:38} & $E,\,a.u.$ & $C = E / Z^{\frac{7}{3}}$& CCSD(T)/cc-pVQZ, $a.u.$ \cite{Kochnev2021}\\
	\hline
	\endfirsthead
	
	\hline
	\multicolumn{6}{| c |}{Continuation of Tab. \ref{E_table_T_0_K}}\\
	\hline
	$Z$ & Element & $\mu,\,a.u.$, \eqref{eq:38} & $E = \mu Z,\,a.u.$ & $C = E / Z^{\frac{7}{3}}$ & CCSD(T)/cc-pVQZ \\
	\hline
	\endhead
	
	\hline
	\endfoot
	
	\hline
	\multicolumn{6}{| c |}{The tolerances $10^{-8}$ for $\mathcal{X}(r)$ and $10^{-9}$ for $\Big|\Big|\frac{\partial I_4}{\partial \mathcal{X}_i}\Big|\Big|_\infty$ are set, and $r_{max} = 470\, a.u.$ }\\
	\hline\hline
	\endlastfoot
2 & He & $-1.614266082$ & $-3.228532163$ & $-0.640622$ & not applicable \\
3 & Li & $-2.771808682$ & $-8.315426046$ & $-0.640622$ & $-7.43272$ \\
4 & Be & $-4.067692597$ & $-16.27077039$ & $-0.640621$ & $-14.61895$ \\
5 & N & $-5.477235428$ & $-27.38617714$ & $-0.640621$ & $-24.60084$ \\
6 & C & $-6.984516663$ & $-41.90709998$ & $-0.640621$  & $-37.73470$ \\
7 & N & $-8.578250255$ & $-60.04775179$ & $-0.640621$ & $-54.42604$ \\
8 & O & $-10.24993906$ & $-81.9995125$ & $-0.640621$ & $-74.91324$ \\
9 & F & $-11.9929115$ & $-107.9362035$ & $-0.640621$ & $-99.65026$ \\
10 & Ne & $-13.80176414$ & $-138.0176414$ & $-0.640621$ & $-128.84369$ \\
11 & Na & $-15.6720149$ & $-172.3921639$ & $-0.640621$ & $-161.85872$ \\
12 & Mg & $-17.5998746$ & $-211.1984952$ & $-0.640621$ & $-199.64815$ \\
13 & Al & $-19.58208988$ & $-254.5671685$ & $-0.640621$ & $-241.93367$ \\
14 & Si & $-21.61583136$ & $-302.621639$ & $-0.640621$ & $-288.90621$ \\
15 & P & $-23.69861151$ & $-355.4791726$ & $-0.640621$ & $-340.76712$ \\
16 & S & $-25.828223$ & $-413.2515679$ & $-0.640621$ & $-397.61899$ \\
17 & Cl & $-28.00269138$ & $-476.0457534$ & $-0.640621$ & $-459.69331$ \\
18 & Ar & $-30.22023806$ & $-543.964285$ & $-0.640621$ & $-526.07291$ \\
19 & K & $-32.47925096$ & $-617.1057683$ & $-0.640621$ & not convergent \\
20 & Cu & $-34.77826089$ & $-695.5652178$ & $-0.640621$ & $-676.78700$ \\
21 & Sc & $-37.11592227$ & $-779.4343676$ & $-0.640621$ & $-759.78204$ \\
22 & Ti & $-39.49099732$ & $-868.801941$ & $-0.640621$ &  $-848.42225$ \\
23 & V & $-41.90234284$ & $-963.7538852$ & $-0.640621$ & $-942.90519$ \\
24 & Cr & $-44.34889918$ & $-1064.37358$ & $-0.640621$ &not convergent \\
25 & Mn & $-46.82968083$ & $-1170.742021$ & $-0.640621$ &not convergent\\
26 & Fe & $-49.34376855$ & $-1282.937982$ & $-0.640621$ &not convergent\\
27 & Co & $-51.89030235$ & $-1401.038164$ & $-0.640621$ &not convergent\\
28 & Ni & $-54.46847578$ & $-1525.117322$ & $-0.640621$ & $-1507.17727$ \\
29 & Cu & $-57.07753067$ & $-1655.248389$ & $-0.640621$ & $-1639.39747$ \\
30 & Zn & $-59.71675272$ & $-1791.502582$ & $-0.640621$ & $-1778.37342$ \\
31 & Ga & $-62.38546756$ & $-1933.949494$ & $-0.640621$ & $-1923.45162$ \\
32 & Ge & $-65.08303728$ & $-2082.657193$ & $-0.640621$ & $-2075.52816$ \\
33 & As & $-67.80885741$ & $-2237.692295$ & $-0.640621$ & $-2234.39397$ \\
34 & Se & $-70.56235412$ & $-2399.12004$ & $-0.640621 $& $-2400.07647$ \\
35 & Br & $-73.34298197$ & $-2567.004369$ & $-0.640621$ & $-2572.73334$ \\
36 & Kr & $-76.15022158$ & $-2741.407977$ & $-0.640621 $& $-2752.38213$ \\
37 & Rb & $-78.98357753$ & $-2922.392369$ & $-0.640621$ & unavailable\\
38 & Sr & $-81.84257722$ & $-3110.017934$ & $-0.640621$ & --<<-- \\
39 & Y & $-84.72676868$ & $-3304.343979$ & $-0.640621$ &--<<--\\
40 & Zn & $-87.63571919$ & $-3505.428768$ & $-0.640621$ &--<<--\\
41 & Nb & $-90.56901437$ & $-3713.329589$ & $-0.640621$ &--<<--\\
42 & Mo & $-93.52625658$ & $-3928.102776$ & $-0.640621$ &--<<--\\
43 & Tc & $-96.50706437$ & $-4149.803768$ & $-0.640621$ &--<<--\\
44 & Ru & $-99.51107057$ & $-4378.487105$ & $-0.640621$ &--<<--\\
45 & Rh & $-102.5379226$ & $-4614.206516$ & $-0.640621$ &--<<--\\
46 & Pd & $-105.5872805$ & $-4857.014902$ & $-0.640621$ &--<<--\\
47 & Ag & $-108.6588169$ & $-5106.964394$ & $-0.640621$ &--<<--\\
48 & Cd & $-111.7522161$ & $-5364.106375$ & $-0.640621$ &--<<--\\
49 & In & $-114.8671732$ & $-5628.491489$ & $-0.640621$ &--<<--\\
50 & Sn & $-118.0033941$ & $-5900.169704$ & $-0.640621$ &--<<--\\
51 & Sb & $-121.1605941$ & $-6179.190301$ & $-0.640621$ &--<<--\\
52 & Te & $-124.338498$ & $-6465.601899$ & $-0.640621$ &--<<--\\
53 & I & $-127.5368397$ & $-6759.452504$ & $-0.640621$ &unavailable\\
54 & Xe & $-130.7553612$ & $-7060.789503$ & $-0.640621$ &--<<--\\
55 & Cs & $-133.9938124$ & $-7369.659681$ & $-0.640621$ &--<<--\\
56 & Ba & $-137.2519509$ & $-7686.109253$ & $-0.640621$ &--<<--\\
57 & La & $-140.529542$ & $-8010.183895$ & $-0.640621$ &--<<--\\
58 & Ce & $-143.8263573$ & $-8341.928721$ & $-0.640621 $&--<<--\\
59 & Pr & $-147.1421751$ & $-8681.388333$ & $-0.640621$ &--<<--\\
60 & Nd & $-150.47678$ & $-9028.606803$ & $-0.640621$ &--<<--\\
61 & Pm & $-153.8299629$ & $-9383.627736$ & $-0.640621$ &--<<--\\
62 & Sm & $-157.20152$ & $-9746.494239$ & $-0.640621$ &--<<--\\
63 & Eu & $-160.5912533$ & $-10117.24896$ & $-0.640621$ &--<<--\\
64 & Gd & $-163.9989698$ & $-10495.93407$ & $-0.640621$ &--<<--\\
65 & Tb & $-167.4244816$ & $-10882.5913$ & $-0.640621$ &--<<--\\
66 & Dy & $-170.867606$ & $-11277.26200$ & $-0.640621$ &--<<--\\
67 & Ho & $-174.3281644$& $-11679.98702$ & $-0.640621$ &--<<--\\
68 & Er & $-177.8059828$ & $-12090.80683$ & $-0.640621$ &--<<--\\
69 & Tm & $-181.300892$ & $-12509.76155$ & $-0.640621$ &--<<--\\
70 & Yb & $-184.812726$ & $-12936.89082$ & $-0.640621$ &--<<--\\
71 & Lu & $-188.3413234$ & $-13372.23396$ & $-0.640621$ &--<<--\\
72 & Hf & $-191.8865263$ & $-13815.82989$ & $-0.640621$ &--<<--\\
73 & Ta & $-195.4481808$ & $-14267.71720$ & $-0.640621$ &--<<--\\
74 & W & $-199.0261361$ & $-14727.93407$ & $-0.640621$ &--<<--\\
75 & Re & $-202.620245$ & $-15196.51837$ & $-0.640621$ &--<<--\\
76 & Os & $-206.2303638$ & $-15673.50765$ & $-0.640621$ &--<<--\\
77 & Ir & $-209.8563516$ & $-16158.93907$ & $-0.640621$ &--<<--\\
78 & Pt & $-213.4980706$ & $-16652.84951$ & $-0.640621$ &--<<--\\
79 & Au & $-217.1553862$ & $-17155.27551$ & $-0.640621$ &unavailable\\
80 & Hg &$-220.8281665$ & $-17666.25332$ & $-0.640621$ &--<<--\\
81 & Tl & $-224.5162824$ & $-18185.81887$ & $-0.640621 $& --<<--\\
82 & Pb & $-228.2196072$ & $-18714.00779$ & $-0.640621$ & --<<--\\
83 & Bi & $-231.9380171$ & $-19250.85542$ & $-0.640621$ & --<<--\\
84 & Po & $-235.6713907$ & $-19796.39682$ & $-0.640621$ & --<<--\\
85 & At & $-239.4196091$ & $-20350.66678$ & $-0.640621$ & --<<--\\
86 & Rn & $-243.1825555$ & $-20913.69978$ & $-0.640621$ & --<<--\\
87 & Fr & $-246.9601155$ & $-21485.53005$ & $-0.640621 $& --<<--\\
88 & Ra & $-250.752177$ & $-22066.19158$ & $-0.640621$ & --<<--\\
89 & Ac & $-254.5586298$ & $-22655.71805$ & $-0.640621$ & --<<--\\
90 & Tr & $-258.379366$ & $-23254.14294$ & $-0.640621$ & --<<--\\
91 & Pa & $-262.2142795$ & $-23861.49943$ & $-0.640621$ & --<<--\\
92 & U & $-266.0632665$ & $-24477.82051$ & $-0.640621$ & --<<--\\
93 & Np & $-269.9262245$ & $-25103.13887$ & $-0.640621$ & --<<--\\
94 & Pu & $-273.8030532$ & $-25737.48700$ & $-0.6406201$ & --<<--\\
95 & Am & $-277.6936541$ & $-26380.89714$ & $-0.640621$ & --<<--\\
96 & Cm & $-281.5979307$ & $-27033.40134$ & $-0.640621$ & --<<--\\
97 & Bk & $-285.5157873$ & $-27695.03137$ & $-0.640621$ & unavailable\\
98 & Cf & $-289.4471309$ & $-28365.81883$ & $-0.640621$ & --<<--\\
99 & Es & $-293.3918693$ & $-29045.79506$ & $-0.640621$ & --<<--\\
100 & Fm & $-297.3499122$ & $-29734.99122$ & $-0.640621$ & --<<--\\
101 & Md & $-301.3211706$ & $-30433.43823$ & $-0.640621$ & --<<--\\
102 & No & $-305.3055575$ & $-31141.16687$ & $-0.640621$ & --<<--\\
103 & Lr & $-309.3029867$ & $-31858.20763$ & $-0.640621$ & --<<--\\
104 & Rf & $-313.3133739$ & $-32584.59088$ & $-0.640621$ & --<<--\\
105 & Df & $-317.3366354$ & $-33320.34672$ & $-0.640621$ & --<<--\\
106 & Sg & $-321.3726898$ & $-34065.50511$ & $-0.640621$ & --<<--\\
107 & Bh & $-325.4214562$ & $-34820.09581$ & $-0.640621$ & --<<--\\
108 & Hs & $-329.4828554$ & $-35584.14839$ & $-0.640621$ & --<<--\\
109 & Mt & $-333.5568094$ & $-36357.69222$ & $-0.640621$ & --<<--\\
110 & Ds & $-337.6432411$ & $-37140.75652$ & $-0.640621$ & --<<--\\
111 & Rg & $-341.7420749$ & $-37933.37031$ & $-0.640621$ & --<<--\\
112 & Cn & $-345.8532361$ & $-38735.56244$ & $-0.640621$ & --<<--\\
113 & Nh & $-349.9766513$ & $-39547.36160$ & $-0.640621$ & --<<--\\
114 & Fl & $-354.1122481$ & $-40368.79628$ & $-0.640621$ & --<<--\\
115 & Mc & $-358.2599548$ & $-41199.89481$ & $-0.640621$ & --<<--\\
116 & Lv & $-362.4197019$ & $-42040.68542$ & $-0.640621$ & --<<--\\
117 & Ts & $-366.5914194$ & $-42891.19606$ & $-0.640621$ & --<<--\\
118 & Og & $-370.7750391$ & $-43751.45462$ & $-0.640621$ & unavailable\\
119 & Uue & $-374.9704939$ & $-44621.48877$ & $-0.640621$ & --<<--\\
120 & Ubn & $-379.1777172$ & $-45501.32606$ & $-0.640621$ & --<<--\\
121 & Ubu & $-383.3966434$ & $-46390.99385$ & $-0.640621$ & --<<--\\
122 & Ubb & $-387.6272085$ & $-47290.51944$ & $-0.640621$ & --<<--\\
123 & Ubt & $-391.8693483$ & $-48199.92984$ & $-0.640621$ & --<<--\\
124 & Ubq & $-396.1230000$ & $-49119.25200$ & $-0.640621$ & --<<--\\
125 & Ubp & $-400.3881017$ & $-50048.51271$ & $-0.640621$ & --<<--\\
126 & Ubh & $-404.6645923$ & $-50987.73863$ & $-0.640621$ & --<<--\\
127 & Ubs & $-408.9524117$ & $-51936.95629$ & $-0.640621$ & --<<--\\
\end{longtable}

\newpage

\fancyhf{}
\renewcommand{\refname}{References}

\end{document}